\documentclass{aa}

\usepackage{hyperref}
\usepackage{graphicx,natbib,bm,color}
\graphicspath{{./fig/}{./png/}}

\newcommand{\EQ}{\begin{equation}}
\newcommand{\EN}{\end{equation}}
\newcommand{\EQA}{\begin{eqnarray}}
\newcommand{\ENA}{\end{eqnarray}}

\newcommand{\EEq}[1]{Equation~(\ref{#1})}
\newcommand{\Eq}[1]{Eq.~(\ref{#1})}
\newcommand{\Eqs}[2]{Eqs.~(\ref{#1}) and~(\ref{#2})}

\newcommand{\App}[1]{Appendix~\ref{#1}}

\newcommand{\Sec}[1]{Sect.~\ref{#1}}

\newcommand{\Fig}[1]{Fig.~\ref{#1}}
\newcommand{\FFig}[1]{Figure~\ref{#1}}

\newcommand{\Tab}[1]{Table~\ref{#1}}

\newcommand{\bra}[1]{\langle #1\rangle}

{}
{}

{}
{}

{}
{}
{}
{}
{}
{}
{}
{}
{}
{}
{}
{}
{}
{}

{}

{}

{}
{}
{}

\newcommand{\zzz}{\hat{\mbox{\boldmath $z$}} {}}

\newcommand{\BB}{\bm{B}}

\newcommand{\JJ}{\bm{J}}

\newcommand{\AAA}{\bm{A}}

\newcommand{\uu}{\bm{u}}

\newcommand{\nab}{{\bm{\nabla}}}

\newcommand{\SSSS}{\mbox{\boldmath ${\sf S}$} {}}

\newcommand{\DD}{{\rm D} {}}

\newcommand{\dd}{{\rm d} {}}

\newcommand{\const}{{\rm const}  {}}

\def\Ma{\mbox{\rm Ma}}

\def\Pm{\mbox{\rm Pr}_{\rm M}}

\def\Rey{\mbox{\rm Re}}

\def\Lu{\mbox{\rm Lu}}

\def\EEK{{\cal E}_{\rm K}}
\def\EEM{{\cal E}_{\rm M}}
\def\EEMz{{\cal E}_{\rm M0}}

\def\EK{E_{\rm K}}
\def\EM{E_{\rm M}}

\def\cs{c_{\rm s}}

\def\xiM{\xi_{\rm M}}
\def\xiMz{\xi_{\rm M0}}
\def\xiK{\xi_{\rm K}}

\def\vA{v_{\rm A}}

\def\vAz{v_{\rm A0}}

\def\kp{k_{\rm p}}

\def\EM{E_{\rm M}}

\def\epsK{\epsilon_{\rm K}}
\def\epsM{\epsilon_{\rm M}}

\def\Brms{B_{\rm rms}}

\def\urms{u_{\rm rms}}

\newcommand{\G}{\,{\rm G}}

\newcommand{\pc}{\,{\rm pc}}

\newcommand{\Mpc}{\,{\rm Mpc}}

\newcommand{\AU}{\,{\rm AU}}

\def\apjl{Astrophys. J. Lett.}
\def\apjs{ApJS}

\def\aap{Astron. Astrophys.}

\titlerunning{}
\authorrunning{Brandenburg et al.}
\title{Resistively controlled primordial magnetic turbulence decay}
\author{A. Brandenburg\inst{1,2,3,4,5},
A. Neronov\inst{6,7}, and
F. Vazza\inst{8,9,10}
}
\institute{
Nordita, KTH Royal Institute of Technology and Stockholm University, Hannes Alfv\'ens v\"ag 12, 10691 Stockholm, Sweden
\and
The Oskar Klein Centre, Department of Astronomy, Stockholm University, AlbaNova, 10691 Stockholm, Sweden
\and
McWilliams Center for Cosmology \& Department of Physics, Carnegie Mellon University, 5000 Forbes Ave, Pittsburgh, PA 15213, USA
\and
School of Natural Sciences and Medicine, Ilia State University, 3-5 Cholokashvili Avenue, 0194 Tbilisi, Georgia
\and
Department of Astrophysics, American Museum of Natural History, 200 Central Park West, New York, NY 10024, USA
\and
Astroparticules et Cosmologie, Universit\'e Paris Cit\'e, CNRS, Astroparticule et Cosmologie, 75006 Paris, France
\and
Laboratory of Astrophysics, \'Ecole Polytechnique F\'ed\'erale de Lausanne, 1015 Lausanne, Switzerland
\and
Dipartimento di Fisica e Astronomia, Universita di Bologna, Via Gobetti 93/2, 40129 Bologna, Italy
\and
INAF Istituto di Radioastronomia, Via P. Gobetti 101, 40129 Bologna, Italy
\and
Hamburger Sternwarte, Universit\"at Hamburg, Gojenbergsweg 112, 41029 Hamburg, Germany
}

\date{\today}
\begin{document}

\abstract{
Magnetic fields generated in the early Universe undergo turbulent decay
during the radiation-dominated era.
The decay is governed by a decay exponent and a decay time.
It has been argued that the latter is prolonged by magnetic reconnection,
which depends on the microphysical resistivity and viscosity.
Turbulence, on the other hand, is not usually expected to be sensitive
to microphysical dissipation, which affects only very small scales.
}{
We want to test and quantify the reconnection hypothesis in decaying
hydromagnetic turbulence.
}{
We performed high-resolution numerical simulations with zero net magnetic
helicity using the \textsc{Pencil Code} with up to $2048^3$ mesh
points and relate the decay time to the Alfv\'en time for different
resistivities and viscosities.
}{
The decay time is found to be longer than the Alfv\'en time by a
factor that increases with increasing Lundquist number to the 1/4 power.
The decay exponent is as expected from the conservation of the Hosking
integral, but a timescale dependence on resistivity is unusual for
developed turbulence and not found for hydrodynamic turbulence.
In two dimensions, the Lundquist number dependence is shown to be leveling
off above values of $\approx25,000$, independently of the value of
the viscosity.
}{
Our numerical results suggest that resistivity effects have been
overestimated in earlier work.
Instead of reconnection, it may be the magnetic helicity density
in smaller patches that is responsible for the resistively slow decay.
The leveling off at large Lundquist number cannot currently be confirmed
in three dimensions.
}

\maketitle

\section{Introduction}

Decaying turbulence played an important role in the early
Universe during the radiation-dominated era, when the magnetic field is
well coupled to the plasma.
While turbulence speeds up the decay, it can also lead to a significant
increase in the typical length scale, which could then be many times
larger than the comoving horizon scale at the time of magnetic field
generation \citep{BEO96,CHB01,BJ04}.
This is important because magnetogenesis processes during the electroweak
era, when the age of the Universe was just a few picoseconds \citep{Vac91,
Cheng+Olinto94, Baym+96}, tend to produce magnetic fields of very small
length scales of the order of $1\AU$ or less.

From the study of decaying hydrodynamic turbulence, it has been known
for a long time that turbulent energy density and length scale evolve
as power laws \citep{Batchelor53, Saffman67}.
The exponents depend on the physics of the decay, specifically on the
possibility of a conserved quantity that governs the decay, for example
magnetic helicity in the hydromagnetic case \citep{Hatori84, BM99}.
The endpoints of the evolution, however, depend on the relevant
timescale, which is traditionally taken to be just the turnover or,
in the magnetic case that we consider here, the Alfv\'en time
\citep{BJ04}.
More recently, \cite{HS23} argued that the turnover time should be
replaced by the reconnection time, which could be significantly longer
(up to $10^{5.5}$ times).
This would result in an endpoint where the magnetic field strength is
greater and the turbulent length scale smaller than otherwise, when the
decay time is just the Alfv\'en time.

One of the hallmarks of turbulence is that its large-scale
properties are nearly independent of viscosity and resistivity, which
act predominantly on the smallest scales of the system.
On the other hand, magnetic reconnection is a process that could
potentially limit the speed of the inverse cascade.
This idea has been invoked by \cite{HS23} to explain a premature
termination of the decay process by the time of recombination of the
Universe, when its age was about 400,000 years.

The purpose of our paper is to analyze numerical simulations with
respect to their decay times at different values of resistivity.
In \Sec{DecayTimes} we discuss the decay time and its relation to
other quantities.
In \Sec{Simulations} we present our numerical simulation setup
and show the results for a resistivity-dependent decay in
\Sec{ResistivityDependentDecay}.
We then make a comparison with a purely hydrodynamic decay in \Sec{hydro} and
with the two-dimensional (2D) hydromagnetic case in \Sec{2d}, before
concluding in \Sec{Conclusions}.
In \App{NoteAnastrophy} we provide a historical note on the
anastrophy, i.e., the mean squared magnetic vector potential, and
in \App{ResolvingPlasmoidInstability}, we show detailed convergence
tests for some of our 2D results.

\section{Decay and turnover times}
\label{DecayTimes}

In the following we focus on the decay of magnetic field.
Magnetically dominated turbulence is characterized by the turbulent
magnetic energy density $\EEM$ and the magnetic integral scale $\xiM$.
Both $\EEM(t)$ and $\xiM(t)$ can be defined in terms of the magnetic
energy spectrum $\EM(k,t),$ such that $\EEM=\int\EM\,\dd k$ and
$\xiM=\int k^{-1}\EM\,\dd k/\EEM$.
In decaying turbulence, both quantities depend algebraically rather than
exponentially on time.
Therefore, the decay is primarily characterized by power laws,
\begin{equation}
\EEM\propto t^{-p}\quad\mbox{and}\quad
\xiM\propto t^{q},
\label{PowerLaws}
\end{equation}
rather than by exponential laws of the type $\EEM\propto e^{-t/\tau}$.
The algebraic decay is mainly a consequence of nonlinearity.
On the other hand, in decaying hydromagnetic turbulence with significant
cross-helicity, for example, the nonlinearity in the induction equation
is reduced and then the decay is indeed no longer algebraic, but closer
to exponential (see \citealt{BO18}).

An obvious difference between algebraic and exponential decays is that
in the former $\EEM(t)$ is characterized by the nondimensional
quantity $p$, while in the latter it is characterized by the dimensionful
quantity $\tau$.
Following \cite{HS23}, a decay time $\tau$ can also be defined for an algebraic
decay and is then given by
\begin{equation}
\tau^{-1}=-\dd\ln\EEM/\dd t.
\end{equation}
In the present case of a power-law decay, this value of $\tau=\tau(t)$
is time-dependent and can be related to the instantaneous decay exponent
\begin{equation}
p(t)=-\dd\ln\EEM/\dd\ln t
\end{equation}
through $\tau=t/p(t)$ (i.e., no new parameter emerges
except for $t$ itself).
However, a useful way of incorporating new information is by relating
$\tau$ to the Alfv\'en time $\tau_\mathrm{A}=\xiM/\vA$ through
\begin{equation}
\tau=C_\mathrm{M}\xiM/\vA,
\label{tau_def}
\end{equation}
where $C_\mathrm{M}$ is a nondimensional parameter, and $\vA$ is the Alfv\'en
velocity, which is related to the magnetic energy density through
$\EEM=\Brms^2/2\mu_0=\rho\vA^2/2$, where $\rho$ is the density,
$\mu_0$ the vacuum permeability, and $\Brms$ the root mean square (rms)
magnetic field.

As was noted by \cite{HS23}, \Eq{tau_def} can be used to define the
endpoints of the evolutionary tracks in a diagram of $\Brms$ versus
$\xiM$ or, equivalently, $\vA$ versus $\xiM$ (i.e., $\vA=\vA(\xi))$.
They also noted that the location of these endpoints is sensitive
to whether or not $C_\mathrm{M}$ depends on the resistivity of the plasma.
If it does depend on the resistivity, this could be ascribed to the effects
of magnetic reconnection, which might slow down the turbulent decay.

Magnetic reconnection refers to a change in magnetic field line
connectivity that is subject to topological constraints.
A standard example is \textsf{x}-point reconnection
\citep[see, e.g.,][]{Craig+McClymont91, Craig+05}, which becomes slower
as the \textsf{x}-point gets degenerated into an extremely elongated
structure \citep{Sweet58,Parker57} (see \citealt{Liu+22} for a review).
It is usually believed that in the presence of turbulence, such structures
break up into progressively smaller ones, which makes reconnection
eventually fast (i.e., independent of the microphysical resistivity)
\citep{Galsgaard+Nordlund96, Lazarian+Vishniac99, Comisso+Sironi19}.
However, whether this would also imply that $\tau$
becomes independent of the resistivity remains an unclear issue.
For example, \cite{Galsgaard+Nordlund96} found that
resistive heating becomes independent of the value of the resistivity.
Another question concerns the speed at which magnetic flux can be processed
through a current sheet \citep{Kowal+09, Loureiro+12}.
Also of interest is the timescale on which the topology of
the magnetic field changes \citep{Lazarian+20}.
These different timescales may not all address the value of
$C_\mathrm{M}$ that relates the decay time to the Alfv\'en time.

In magnetically dominated turbulence, the effect of the resistivity is
quantified by the Lundquist number.
For decaying turbulence, it is time-dependent and defined as
\begin{equation}
\Lu(t)=\vA(t)\,\xiM(t)/\eta.
\end{equation}
This quantity is similar to the magnetic Reynolds number if we replace
$\vA$ by the rms velocity, $\urms$.
Here, however, the plasma is driven by the Lorentz force, so the
Lundquist number is a more direct way of quantifying the resistivity
than the magnetic Reynolds number.
The Alfv\'enic Mach number is defined as $\Ma_\mathrm{A}=\urms/\vA$.

In addition to varying $\eta$, we also vary $\nu$ such that the magnetic
Prandtl number $\Pm=\nu/\eta$ is typically in the range $1\leq\Pm\leq5$.
It is then also convenient to define the Lundquist number based on the
reconnection outflow,
\begin{equation}
\Lu_\nu(t)=\Lu(t)/\sqrt{1+\Pm}
\end{equation}
(see \citealt{HS23} for details).

\begin{table*}[t!]\caption{
Summary of magnetohydrodynamic simulations analyzed in this paper.
The time unit of $\tau_\xi$ and $\tau_\mathcal{E}$ is $[t]=(\cs k_1)^{-1}$.
}\vspace{12pt}\centerline{\begin{tabular}{ccccccccccccccc}
Run & $\eta k_1/\cs$ & $\nu k_1/\cs$ & $\Pm$ & $\Lu$ & $\Lu_\nu$ & $C_\mathrm{M}$ & $C_\mathrm{L}^{(1/4)}$ &
$C_\xi$ & $C_\mathcal{E}$ & $\tau_\xi$ & $\tau_\mathcal{E}$ & $\!\!\epsK/\epsM\!\!$ & $\Ma_\mathrm{A}$ & $N^3$\\
\hline
M0 & $4\times10^{-7}$ & $4\times10^{-6}$ &10 & 5400 & 1600 & $36.0\pm1.4$ & 4.2 & 91 & 0.39 & 0.82 & 0.28 & 0.9 & 0.20 & $2048^3$ \\
M1 & $4\times10^{-7}$ & $2\times10^{-6}$ & 5 & 5830 & 2600 & $30.5\pm0.8$ & 3.5 & 87 & 0.35 & 0.72 & 0.23 & 0.9 & 0.24 & $2048^3$ \\
M2 & $1\times10^{-6}$ & $2\times10^{-6}$ & 2 & 2354 & 1660 & $26.2\pm0.3$ & 3.8 & 81 & 0.32 & 0.63 & 0.20 & 0.6 & 0.28 & $2048^3$ \\
M3 & $1\times10^{-6}$ & $5\times10^{-6}$ & 5 & 2186 &  980 & $25.6\pm0.6$ & 3.7 & 80 & 0.32 & 0.74 & 0.22 & 0.8 & 0.25 & $1024^3$ \\
M4 &$2.5\times10^{-6}$& $5\times10^{-6}$ & 2 &  823 &  580 & $20.1\pm0.5$ & 3.7 & 73 & 0.27 & 0.61 & 0.17 & 0.6 & 0.31 & $1024^3$ \\
M5 & $5\times10^{-6}$ & $5\times10^{-6}$ & 1 &  386 &  386 & $14.5\pm0.7$ & 3.3 & 64 & 0.23 & 0.45 & 0.12 & 0.5 & 0.38 & $1024^3$ \\
\label{TSummary}\end{tabular}}\end{table*}

\begin{figure}[t!]\begin{center}
\includegraphics[width=\columnwidth]{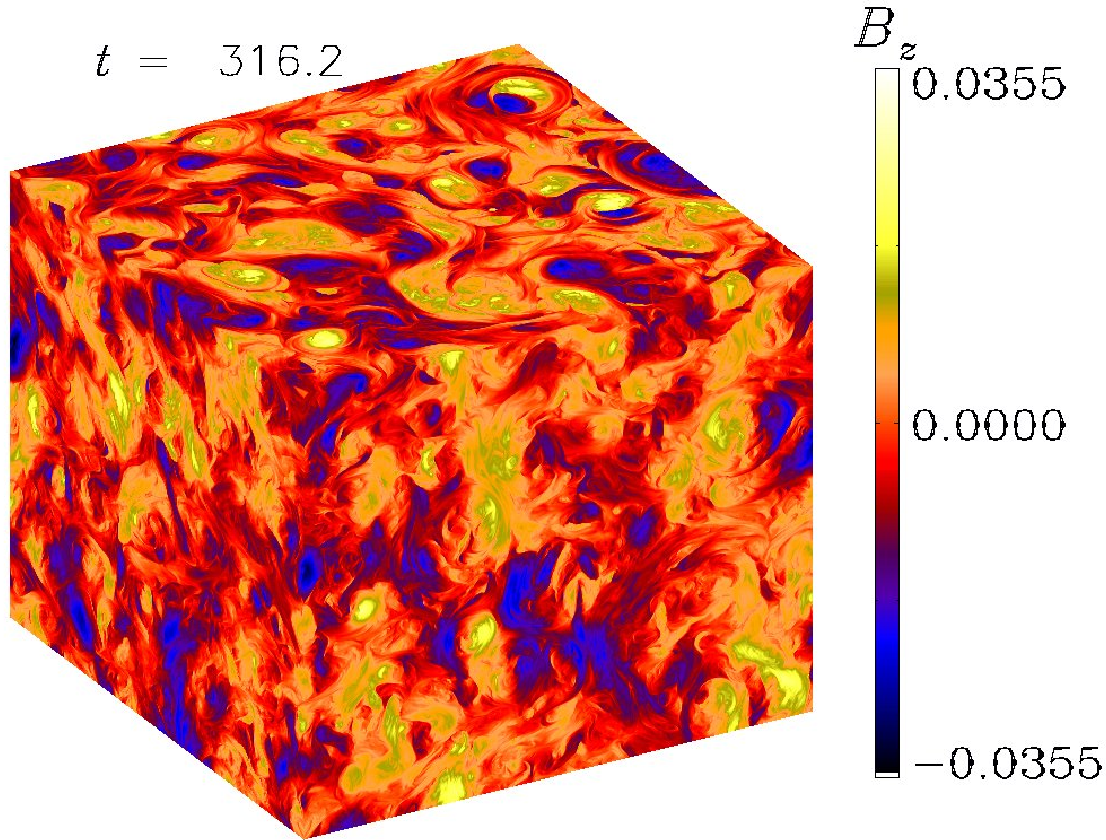}
\end{center}\caption[]{
Visualization of $B_z$ on the periphery of the computational domain
for Run~M1, where $\Pm=5$.
}\label{AB}\end{figure}

\begin{figure}[t!]\begin{center}
\includegraphics[width=\columnwidth]{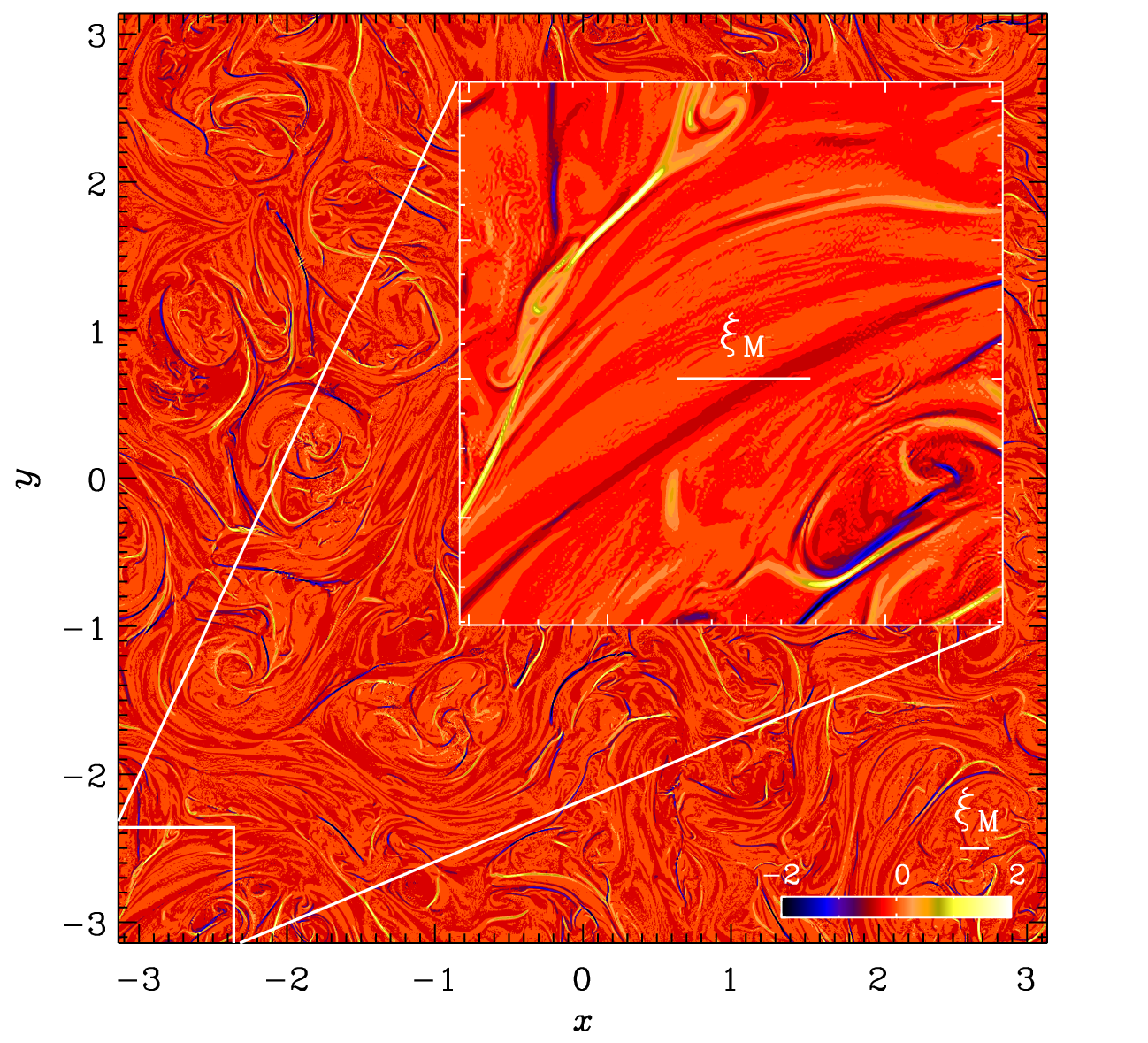}
\end{center}\caption[]{
Visualization of $J_z$ at $z=\pi$ along with a zoom-in on the lower
left corner for Run~M0, where $\Pm=10$, at $t=644$.
The value of $\xiM$ at that time is indicated by the length of the
short white lines.
}\label{pslice}\end{figure}

\section{Numerical simulations}
\label{Simulations}

We performed simulations of the compressible hydromagnetic equations
for the magnetic vector potential $\AAA$, the velocity $\uu$, and the
logarithmic density $\ln\rho$ in the presence of viscosity $\nu$ and
magnetic diffusivity $\eta$: 
\begin{equation}
\frac{\partial\AAA}{\partial t}=\uu\times\BB+\eta\nabla^2\AAA,
\label{dAdt}
\end{equation}
\begin{equation}
\frac{\DD\uu}{\DD t}=
\frac{1}{\rho}\left[\JJ\times\BB+\nab\cdot(2\rho\nu\SSSS)\right]
-\cs^2\nab\ln\rho,
\label{DuDt}
\end{equation}
\begin{equation}
\frac{\DD\ln\rho}{\DD t}=-\nab\cdot\uu.
\end{equation}
We used a random magnetic field as the initial condition, such that $\EM(k,0)$
has a $k^4$ subinertial range for $k<\kp$ \citep{DC03}, and a $k^{-2}$
inertial range for $k>\kp$ \citep{BKT15}.
In all cases we chose $\kp/k_1=60$, where $k_1=2\pi/L$ is the smallest
wavenumber in our cubic domain of size $L^3$.
In \Eqs{dAdt}{DuDt}, $\BB=\nab\times\AAA$ is the magnetic field,
$\JJ=\nab\times\BB/\mu_0$ is the current density, and
${\sf S}_{ij}=(\partial_i u_j+\partial_j u_i)/2-\delta_{ij}\nab\cdot\uu/3$
are the components of the rate-of-strain tensor $\SSSS$.
There is no magnetic helicity on average, but the fluctuations in
the local magnetic helicity density $h=\AAA\cdot\BB$ lead to a decay
behavior where the correlation integral of $h$, which is also known as
the Hosking integral, is conserved \citep{HS21, Scheko22, Zhou+22}.
We use the \textsc{Pencil Code} \citep{JOSS}, which has also been
used for many earlier simulations of decaying hydromagnetic turbulence
\citep{Zhou+22, BSV23}.
All our simulations are in the magnetically dominated regime, because
the velocity is just a consequence of and driven by the magnetic field.

Because the magnetic field is initially random, the resulting velocity
is also random and it drives a forward turbulent cascade with kinetic
and magnetic energy dissipation rates $\epsK=\bra{2\nu\rho\SSSS^2}$
and $\epsM=\bra{\eta\mu_0\JJ^2}$.
Their ratio scales with $\Pm^{1/3}$ \citep{Bra14,Galishnikova+22}.
If $\kp/k_1$ is large (we recall that we use the value 60),
there is also an inverse cascade \citep{BKT15}.
The inverse cascade was also found in the relativistic regime
\citep{Zrake14} and is now understood to be a consequence of the
conservation of the Hosking integral \citep{HS21, Scheko22, Zhou+22}.
However, the role played by the Hosking integral is currently not
universally accepted \citep{Armua+23, Dwivedi+24}.
The lack of numerical support could be related to insufficiently large
values of $\kp/k_1$.

\begin{figure*}[t!]\begin{center}
\includegraphics[width=\textwidth]{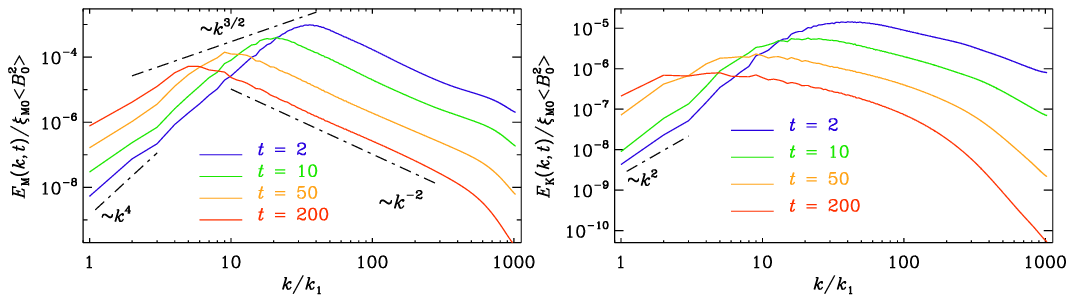}
\end{center}\caption[]{
Magnetic and kinetic energy spectra for Run~M1 at different times.
For large values of $k$ there is a $k^{-2}$ inertial range.
The $\propto k^4$ and $\propto k^2$ slopes are indicated for reference.
As elsewhere, time is in units of $[t]=(\cs k_1)^{-1}$.
}\label{rspec_select_k60del2bc_k4_Pm5}\end{figure*}

We define the kinetic energy spectrum $\EK(k,t)$ analogously to $\EM(k,t)$,
using the normalization $\int\EK(k,t)\,\dd k=\EEK(t)\equiv\rho_0\urms^2/2$,
where $\rho_0=\bra{\rho}=\const$ owing to mass conservation.
In magnetically driven turbulence, $\EEK$ is about one-tenth of
$\EEM=\Brms^2/2\mu_0$, which seems to be surprisingly independent
of the physical input parameters \citep{Bran+17}.

In \Tab{TSummary} we summarize the results of five simulations,
Runs~M0--M5, where we vary $\eta$ and $\nu$ and also vary the number
of mesh points, $N^3$.
We present some relevant output parameters that are defined below.
They are all obtained from a statistically steady stretch of our time
series data, and the error bars were calculated as the largest departure
from any one-third of the time series.
A visualization of the $z$ components of $\BB$ on the periphery
of the computational domain for Run~M1 is shown in \Fig{AB}.
\FFig{pslice} shows $J_z$ in an $xy$ plane together with a zoom-in
on the lower left corner of the domain.
The corresponding magnetic and kinetic energy spectra are shown in
\Fig{rspec_select_k60del2bc_k4_Pm5} for different times in units
of $[t]=(\cs k_1)^{-1}$.

The $\EM(k,t)\propto k^{-2}$ inertial range can be explained
by weak turbulence scaling \citep{BKT15}, but it becomes a bit
shallower near the dissipation subrange.
This could follow from some kind of magnetic bottleneck
effect associated with reconnection, analogously to the bottleneck
effect in hydrodynamic turbulence \citep{Falk94}.

It should be noted that at late times, the subinertial range spectrum of $\EM(k,t)$
becomes shallower than the initial $k^4$ slope.
This is an artifact of poor scale separation (see \citealt{BSV23} for
related numerical evidence).
On the other hand, a $k^2$ subinertial range for $\EK(k,t)$
has been seen for some time (see \citealt{Kahniashvili+13}).

In our numerical simulations we use units such that
$\cs=k_1=\rho_0=\mu_0=1$.
The resistivity, $\mu_0\eta$, is therefore the same as the magnetic diffusivity.
Nevertheless, most of the results below are expressed in manifestly
nondimensional form.

\section{Resistivity-dependent decay}
\label{ResistivityDependentDecay}

We now analyze a collection of runs similar to those of the recent
works of \cite{Zhou+22} and \cite{BSV23}, who considered different
values of $\Lu$ and also included some runs with hyperviscosity and
hyperresistivity, unlike what we present in the present work.
At variance with those earlier papers, where the focus was always on the
decay exponent $p(t)$, here we focus on the evolution of the decay time,
$\tau(t)=t/p(t)$.

\subsection{Decay time}

The goal is to determine the prefactor in the scaling relation $\tau\propto\xiM/\vA$.
Therefore, we write
\begin{equation}
t/p\equiv\tau=C_\mathrm{M}\xiM/\vA
\label{tp_vs_xivA}
\end{equation}
and determine
\begin{equation}
C_\mathrm{M}(t)=(t/p)\,\vA(t)/\xiM(t).
\label{Ctau}
\end{equation}
We emphasize here that each of the terms is time-dependent, including
$\tau(t)=t/p(t)$, as already noted above. 
Interestingly, it turns out that the quantity $C_\mathrm{M}(t)$ eventually
settles around a plateau:
\begin{equation}
C_\mathrm{M}=\lim_{t\to\infty}C_\mathrm{M}(t).
\end{equation}
Here and in the following, when time-dependence is not indicated,
we usually mean that the value is obtained as a suitable limit of the
corresponding time-dependent function.
In numerical simulations with finite domains, the limit
$t\to\infty$ needs to be evaluated with some care to prevent
the final result from being contaminated by finite size effects.
We do this by selecting a suitable time interval during which certain
data combinations are approximately statistically stationary.
We refer to these results as late-time limits.

\begin{figure}[t!]\begin{center}
\includegraphics[width=\columnwidth]{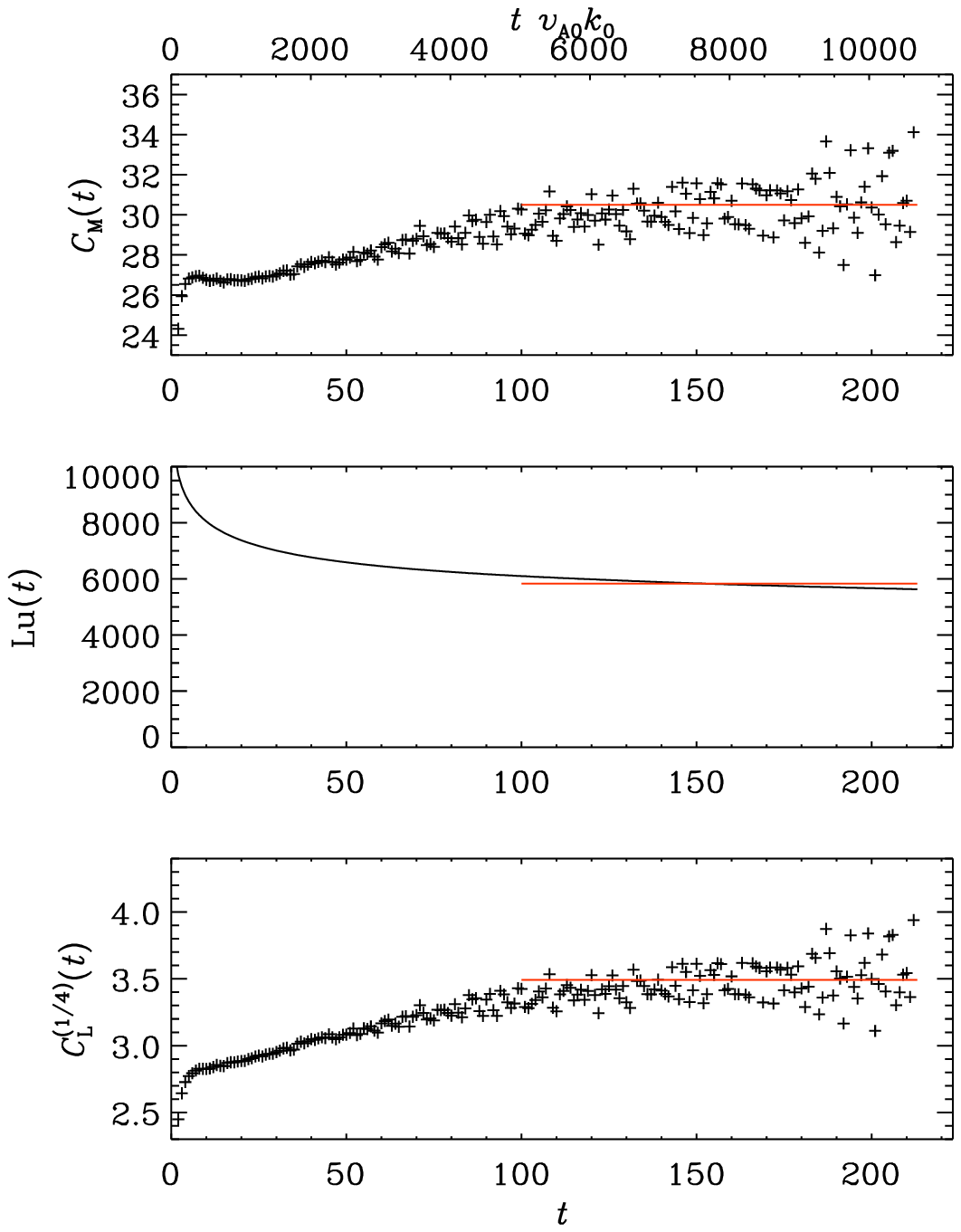}
\end{center}\caption[]{
Approach of $C_\mathrm{M}(t)$, $\Lu(t)$, and $C_\mathrm{L}^{(1/4)}(t)$ to an
approximately constant value toward the end of the simulation for Run~M1.
}\label{pkpm_k60del2bc_k4_Pm5}\end{figure}

In \Fig{pkpm_k60del2bc_k4_Pm5} we show that $C_\mathrm{M}(t)$ and
$\Lu(t)$ approach an approximately constant value toward the end of
the run.
Time is given both in units of $[t]=(\cs k_1)^{-1}$ and in initial
Alfv\'en times, $(\vAz\kp)^{-1}$, where $\vAz=\sqrt{2\EEMz}$ is the
initial Alfv\'en velocity and $\EEMz$ the initial magnetic energy density.
Allowing for the possibility of power-law scaling,
$C_\mathrm{M}=C_\mathrm{L}^{(n)}\Lu^n$, we also plot the prefactor
$C_\mathrm{L}^{(n)}(t)$ for $n=1/4$ in the last panel of
\Fig{pkpm_k60del2bc_k4_Pm5} (see details below).
Toward the end of the simulation, however, there is an increase in the
fluctuations, which follows from the decrease in the magnetic energy.

\begin{figure}[t!]\begin{center}
\includegraphics[width=\columnwidth]{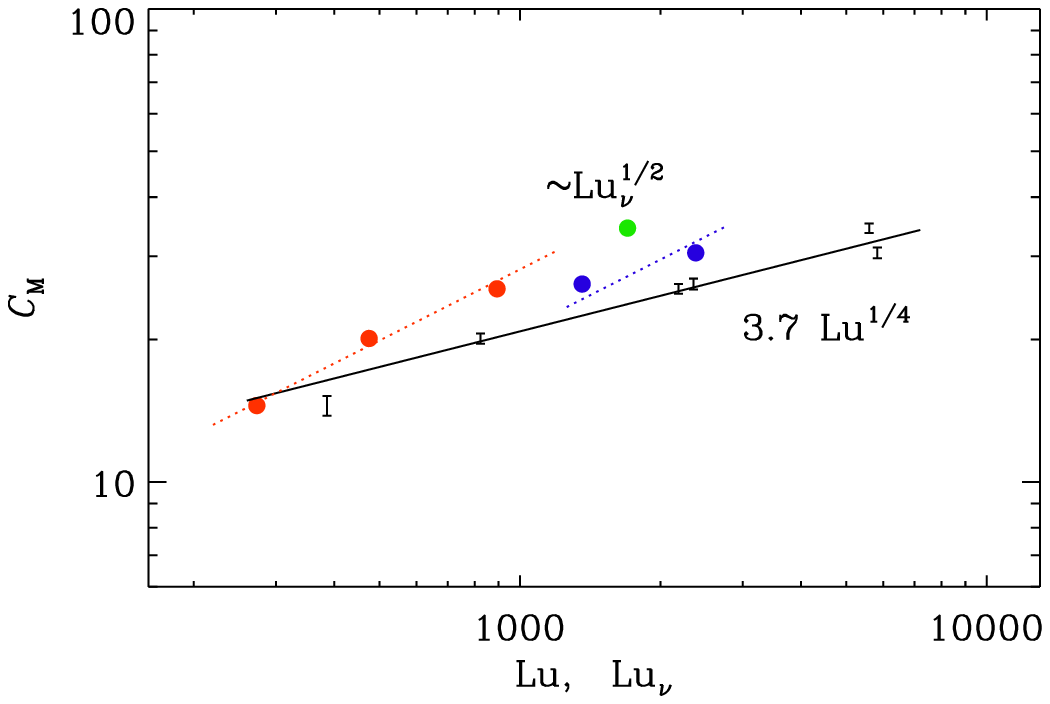}
\end{center}\caption[]{
Dependence of $C_\mathrm{M}$ on $\Lu$ and $\Lu_\nu$.
An approximate scaling $\propto\Lu^{1/4}$ is found for $\Lu<6000$
and a piecewise power-law scaling $\propto\Lu_\nu^{1/2}$ with different
prefactors for the runs with higher viscosity ($5\times10^{-6}$ red
symbols and $4\times10^{-6}$ for the green symbol) and lower viscosity
($2\times10^{-6}$ blue symbols).
}\label{ptab2}\end{figure}

As discussed below in more detail in connection with 2D simulations,
current sheets are underresolved at early times, when $\xiM$ is small.
As we now see from \Fig{pkpm_k60del2bc_k4_Pm5}, this underestimates the
resulting value of $C_\mathrm{M}$.
However, for $t>100$, $C_\mathrm{M}(t)$ approaches a plateau, suggesting
that the simulation now begins to be sufficiently well resolved, at
least for the purpose of determining $C_\mathrm{M}$.

In \Fig{ptab2} we show the dependence of $C_\mathrm{M}$ on $\Lu$
and $\Lu_\nu$.
We see an approximate scaling $\propto\Lu^n$ with $n=1/4$ for $\Lu<6000$
and a piecewise power-law scaling $\propto\Lu_\nu^{1/2}$, but with
different prefactors for the runs with larger and smaller values of
the viscosity.
We note that for $\Pm\gg1$ we have $\Lu_\nu=\vA\xiM/\sqrt{\eta\nu}$, which
explains why the $\Lu_\nu^{1/2}$ scaling is found to be compatible with
being $\propto\Lu^{1/4}$.
We also note, however, that for a higher viscosity, the line $\Lu_\nu^{1/2}$
is shifted upward (toward larger values of $C_\mathrm{M}$).
Owing to the more complicated combined dependence on $\nu$ and $\Lu_\nu$,
we continue to employ the simpler $\Lu^n$ scaling for the following
discussion.
It is worth noting that the exponent $n=1/4$ is discussed in
\cite{Uzdensky+Loureiro16} in connection with the fast growing mode of
the tearing instability.

If it is really true that $C_\mathrm{M}$ is proportional to $\Lu^n$,
as found above, we can write $C_\mathrm{M}=C_\mathrm{L}^{(n)}\Lu^n$, and
then determine $C_\mathrm{L}^{(n)}$ as the late-time limit of
\begin{equation}
C_\mathrm{L}^{(n)}(t)=\frac{t}{p}\,\frac{\vA^{1-n}}{\xiM^{1+n}}\,\eta^n.
\label{CL}
\end{equation}
This is the formula that was used to compute $C_\mathrm{L}^{(n)}$
and to plot its time-dependence $C_\mathrm{L}^{(n)}(t)$ for $n=1/4$
(see the last panel of \Fig{pkpm_k60del2bc_k4_Pm5}).
From \Tab{TSummary} we find $3.3\leq C_\mathrm{L}^{(1/4)}\leq3.8$.

\begin{figure}[t!]\begin{center}
\includegraphics[width=\columnwidth]{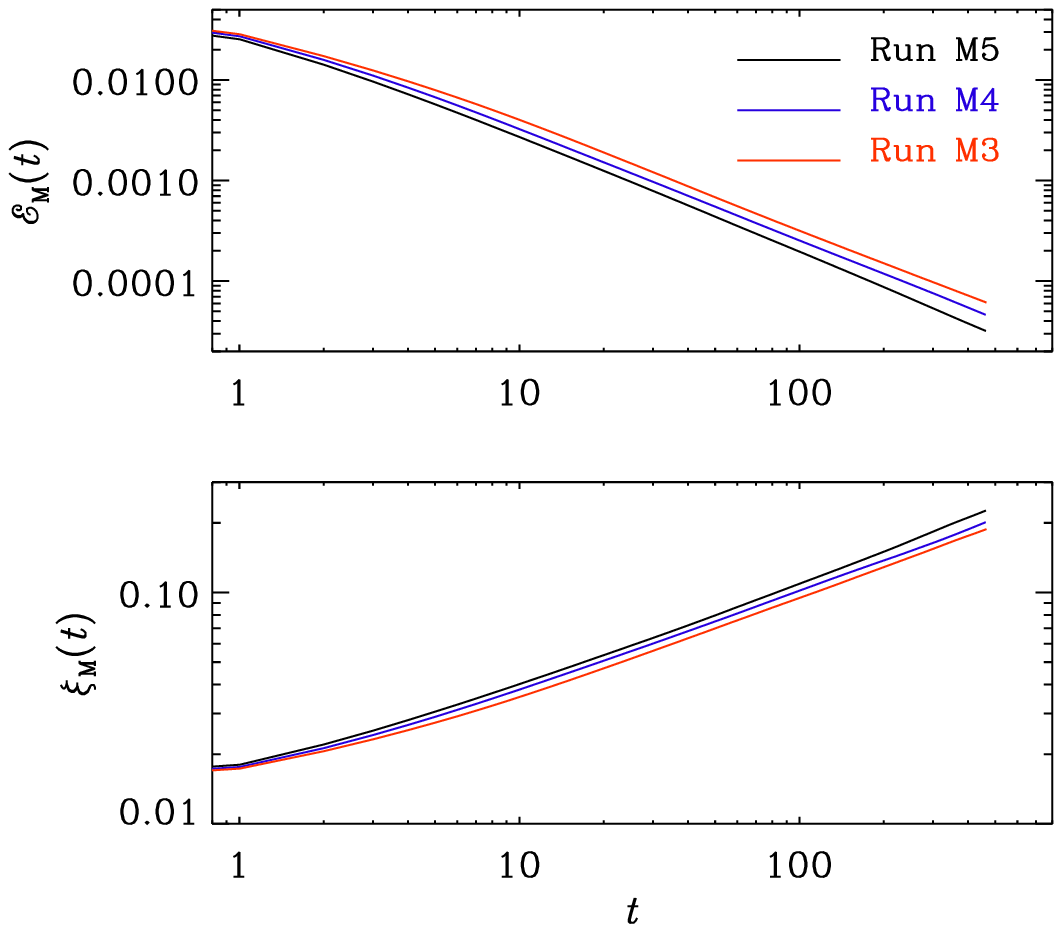}
\end{center}\caption[]{
Dependence of $\EEM(t)$ and $\xiM(t)$ for three values of $\Pm$.
}\label{pEMxi_comp_k60del2bc_nu5em6_k4_Pm1}\end{figure}

\subsection{Evolution of $\EEM(t)$ and $\xiM(t)$}

It is of interest to know whether the
resistivity dependence of $C_\mathrm{M}(t)$ is equally
distributed among $\xiM$ and $\vA$ (or $\EEM$).
\FFig{pEMxi_comp_k60del2bc_nu5em6_k4_Pm1} shows that there are indeed
systematic differences in the decay laws for different values of the
resistivity, but the differences are small and easily overlooked.

\begin{figure}[t!]\begin{center}
\includegraphics[width=\columnwidth]{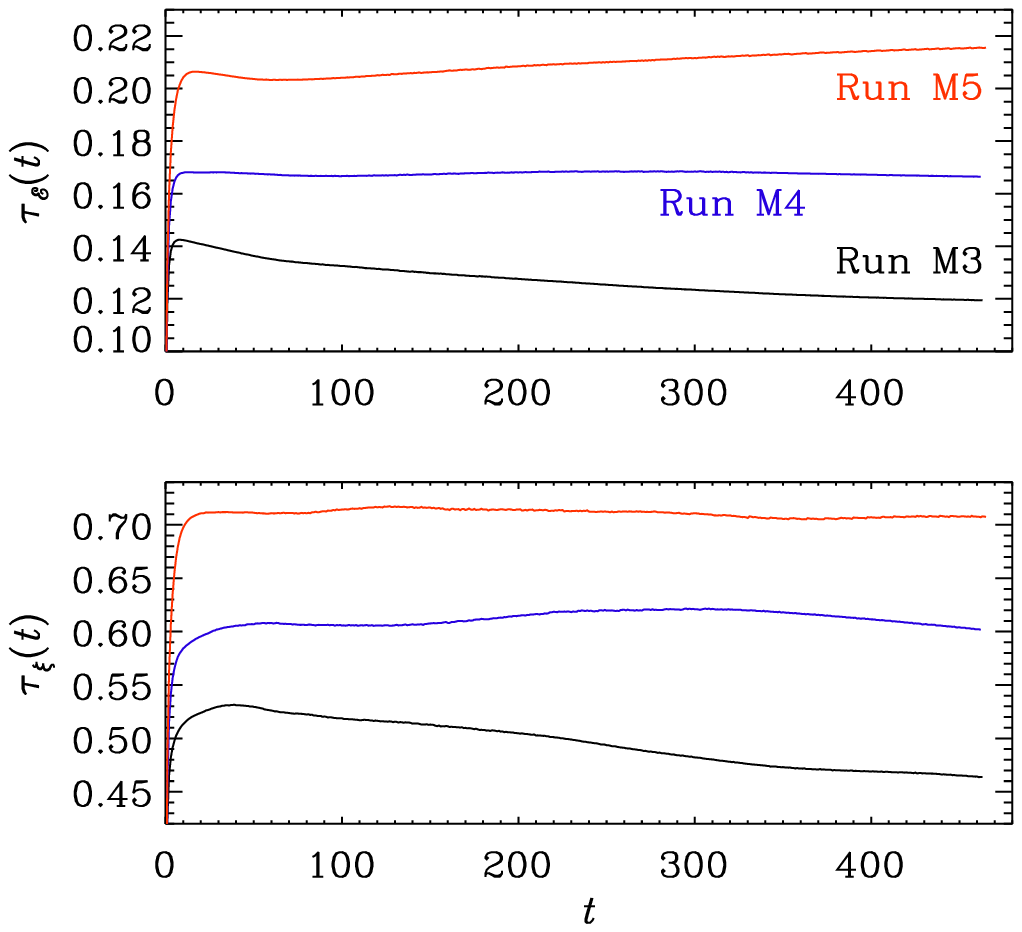}
\end{center}\caption[]{
Evolution of $\tau_\mathcal{E}(t)$ and $\tau_\xi(t)$ for three values
of $\Pm$ with fixed and moderate viscosity.
These times tend to be approximately constant at late times
with values approximately consistent with those in \Tab{TSummary}.
}\label{pEMxi_comp2_k60del2bc_nu5em6_k4_Pm1}\end{figure}

To examine the dependence of $C_\mathrm{M}(t)$ on $\Lu$, we write the
decay laws for $\xiM(t)$ and $\EEM(t)$ in a more detailed form than
\Eq{PowerLaws}: 
\begin{equation}
\xiM(t)=\xiMz\,(1+t/\tau_\xi)^q,
\label{xiMt}
\end{equation}
\begin{equation}
\EEM(t)=\EEMz\,(1+t/\tau_\mathcal{E})^{-p}.
\label{EEMt}
\end{equation}
Here the coefficients $\xiMz$ and $\EEMz$ just depend on the
initial condition, and are thus not dependent on $\Lu$,
which means that the $\Lu$-dependence can only enter through the coefficients
$\tau_\xi$ and $\tau_\mathcal{E}$.
We can determine them as the limits of the time-dependent functions
$\tau_\xi(t)$ and $\tau_\mathcal{E}(t)$, which are obtained by inverting
\Eqs{xiMt}{EEMt}, and are given by
\begin{equation}
\tau_\xi(t)=\frac{t}{\left[\xiM(t)/\xiMz\right]^{1/q}-1},
\label{tauxi}
\end{equation}
\begin{equation}
\tau_\mathcal{E}(t)=\frac{t}{\left[\EEM(t)/\EEMz\right]^{-1/p}-1}.
\label{tauE}
\end{equation}
\FFig{pEMxi_comp2_k60del2bc_nu5em6_k4_Pm1} shows the evolution of
$\tau_\xi(t)$ and $\tau_\mathcal{E}(t)$ for Runs~M3--M5 with fixed and
moderate viscosity.

Using the late-time limits of \Eqs{xiMt}{EEMt}, and those of
\Eqs{tauxi}{tauE}, the equation for
$C_\mathrm{M}$ can now be decomposed in the form
\begin{equation}
C_\mathrm{M}=C_\xi C_\mathcal{E},
\end{equation}
such that \Eq{Ctau} is obeyed.
Here we can determine $C_\xi$ and $C_\mathcal{E}$ as the late-time
limits of
\begin{equation}
C_\xi(t)=\tau^{q}\xiM^{-1}\quad\mbox{and}\quad
C_\mathcal{E}(t)=\tau^{p/2}\vA.
\end{equation}
In this connection, it it important to remember that for a self-similar
evolution \citep{BK17}, which is here approximately satisfied (see
\Fig{rspec_select_k60del2bc_k4_Pm5}), we have
\begin{equation}
q+p/2=1,
\end{equation}
so that $C_\mathrm{M}=(t/p)\,\vA/\xiM$ is obeyed.
For $t\gg\tau_\mathcal{E},\tau_\xi$, using again
$\tau=\tau(t)=t/p(t)$, we have
\begin{equation}
C_\xi\approx(\tau_\xi/p)^{q}\xiMz^{-1}\quad\mbox{and}\quad
C_\mathcal{E}\approx(\tau_\mathcal{E}/p)^{p/2}\vAz,
\label{CxiCEE}
\end{equation}
so that $C_\mathrm{M}=(\vAz/p\xiMz)\,\tau_\xi^q\tau_\mathcal{E}^{p/2}$.
It is then natural to expect that both
$\tau_\xi$ and $\tau_\mathcal{E}$ scale in the same way with $\Lu$
as $C_\mathrm{M}$ itself.
\EEq{CxiCEE} also allows us to compute the timescales
$\tau_\xi=p\,(C_\xi\xiMz)^{1/q}$ and
$\tau_\mathcal{E}=p\,(C_\mathcal{E}/\vAz)^{2/p}$.

\begin{figure}[t!]\begin{center}
\includegraphics[width=\columnwidth]{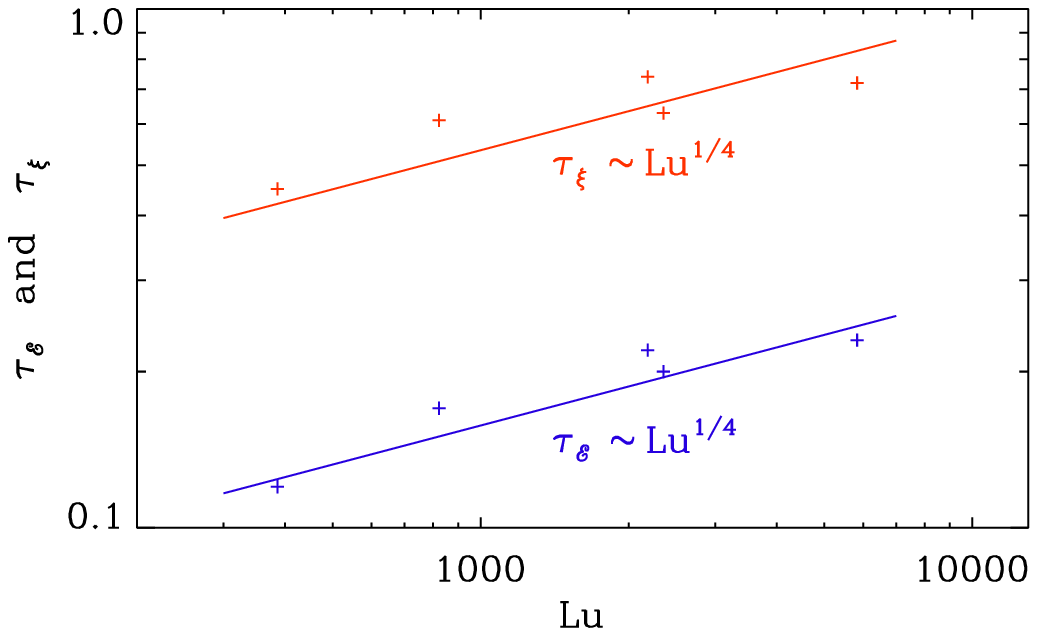}
\end{center}\caption[]{
Dependence of $\tau_\xi$ and $\tau_\mathcal{E}$ on $\Lu$.
The two show a similar dependence on $\Lu$, approximately
$\propto\Lu^{1/4}$, but there is possible evidence for a leveling
off for larger values of $\Lu$.
}\label{ptab_tauExi_all}\end{figure}

\begin{table*}[t]\caption{
Summary of hydrodynamic simulations discussed in \Sec{hydro}.
}\vspace{12pt}\centerline{\begin{tabular}{cccccccccc}
Run & $\nu k_1/\cs$ & $\Rey$ & $C_\mathrm{K}$ & $C_\mathrm{L}^{(1/4)}$ & $C_\xi$ & $C_\mathcal{E}$ & $\tau_\xi$ & $\tau_\mathcal{E}$ & $N^3$\\
\hline
H1 & $1\times10^{-7}$ & 573 & $5.5\pm0.5$ & 1.1 & 230 & 0.024 & 7.5 & 0.12 & $1024^3$ \\
H2 & $2\times10^{-7}$ & 280 & $5.8\pm1.0$ & 1.4 & 234 & 0.024 & 7.8 & 0.12 & $1024^3$ \\
H3 & $5\times10^{-7}$ & 134 & $4.9\pm0.5$ & 1.4 & 205 & 0.024 & 5.8 & 0.12 & $1024^3$ \\
H4 & $1\times10^{-6}$ & 110 & $4.4\pm0.6$ & 1.4 & 164 & 0.027 & 3.5 & 0.16 & $1024^3$ \\
H5 & $2\times10^{-6}$ &  35 & $3.7\pm0.7$ & 1.5 & 167 & 0.022 & 4.9 & 0.13 & $ 512^3$ \\
\label{TSummaryH}\end{tabular}}\end{table*}

\begin{table*}[t!]\caption{
Similar to \Tab{TSummary}, but for the 2D hydromagnetic simulations analyzed in \Sec{2d}.
Here $\kp/k_1=200$.
}\vspace{12pt}\centerline{\begin{tabular}{cccccccccccccc}
Run & $\eta k_1/\cs$ & $\nu k_1/\cs$ & $\Pm$ & $\Lu$ & $C_\mathrm{M}$ & $C_\mathrm{L}^{(1/4)}$ &
$C_\xi$ & $C_\mathcal{E}$ & $\tau_\xi$ & $\tau_\mathcal{E}$ & $\!\!\epsK/\epsM\!\!$ & $\Ma_\mathrm{A}$ & $N^2$\\
\hline
2m1& $5\times10^{-9}$ & $5\times10^{-7}$&100& 75,000 & $36.7\pm4.6$ & 2.2 &221 & 0.164& 0.34 & 1.26 & 13 & 0.41 & $16384^2$\\
2m2& $2\times10^{-9}$ & $2\times10^{-7}$&100&182,000 & $39.5\pm1.0$ & 1.9 &241 & 0.164& 0.41 & 1.25 & 21 & 0.43 & $16384^2$\\
2m3& $1\times10^{-9}$ & $1\times10^{-7}$&100&358,000 & $42.7\pm1.7$ & 1.7 &254 & 0.168& 0.46 & 1.30 & 33 & 0.44 & $16384^2$\\
2m4& $1\times10^{-9}$ & $2\times10^{-8}$ &20&356,000 & $43.8\pm2.5$ & 1.8 &248 & 0.175& 0.50 & 1.50 & 11 & 0.44 & $8192^2$ \\
2m5& $2\times10^{-9}$ & $2\times10^{-8}$ &10&178,000 & $43.9\pm3.0$ & 2.1 &247 & 0.177& 0.49 & 1.53 & 6.0& 0.44 & $8192^2$ \\
2m6& $2\times10^{-9}$ & $2\times10^{-8}$ &10&178,000 & $45.3\pm1.8$ & 2.2 &254 & 0.178& 0.47 & 1.45 & 6.0& 0.44 & $16384^2$\\
2M1& $4\times10^{-9}$ & $2\times10^{-8}$ & 5 &90,800 & $42.5\pm2.1$ & 2.4 &241 & 0.175& 0.47 & 1.34 & 3.5& 0.45 & $8192^2$ \\
2M2& $2\times10^{-8}$ & $1\times10^{-7}$ & 5 &19,000 & $36.8\pm2.6$ & 3.1 &217 & 0.169& 0.37 & 1.40 & 2.9& 0.45 & $8192^2$ \\
2M3& $1\times10^{-7}$ & $5\times10^{-7}$ & 5 & 3900  & $30.2\pm2.0$ & 3.8 &191 & 0.157& 0.34 & 1.36 & 1.5& 0.42 & $4096^2$ \\
2M4& $1\times10^{-7}$ & $5\times10^{-7}$ & 5 &  770  & $18.9\pm0.6$ & 3.6 &347 & 0.054& 1.30 & 3.66 & 0.8& 0.42 & $4096^2$ \\
2M5& $1\times10^{-9}$ & $1\times10^{-9}$ & 1&365,000 & $44.1\pm1.8$ & 1.8 &252 & 0.175& 0.52 & 1.49 & 0.7& 0.44 & $8192^2$ \\
2M6& $5\times10^{-9}$ & $5\times10^{-9}$ & 1 &72,300 & $39.9\pm1.9$ & 2.4 &232 & 0.171& 0.43 & 1.43 & 0.8& 0.46 & $8192^2$ \\
2M7& $2\times10^{-8}$ & $2\times10^{-8}$ & 1 &18,700 & $36.8\pm1.7$ & 3.1 &233 & 0.157& 0.44 & 1.23 & 0.8& 0.46 & $8192^2$ \\
2M8& $1\times10^{-7}$ & $1\times10^{-7}$ & 1 & 3830  & $31.0\pm0.7$ & 3.9 &213 & 0.146& 0.35 & 1.07 & 0.7& 0.45 & $8192^2$ \\
2M9& $5\times10^{-7}$ & $5\times10^{-7}$ & 1 &  742  & $18.7\pm0.8$ & 3.6 &156 & 0.119& 0.22 & 0.83 & 0.5& 0.51 & $4096^2$ \\
2M10&$5\times10^{-7}$ & $5\times10^{-7}$ & 1 &  125  & $10.0\pm0.4$ & 3.0 &284 & 0.035& 0.83 & 1.66 & 0.5& 0.63 & $4096^2$ \\
\label{TSummary2D}\end{tabular}}\end{table*}

In \Fig{ptab_tauExi_all} we show the dependence of $\tau_\xi$ and
$\tau_\mathcal{E}$ on $\Lu$.
While the two show a similar dependence, approximately
$\propto\Lu^{1/4}$, we note that there is also possible evidence for a leveling off
for larger values of $\Lu$.

\section{Comparison with hydrodynamic decay}
\label{hydro}

Hydrodynamic decay is characterized by the kinetic energy
density $\EEK(t)=\rho_0\urms^2/2$ and the hydrodynamic integral
scale $\xiK(t)=\int k^{-1}\EK\,\dd k/\EEK$.
We define the instantaneous kinetic energy decay exponent
$p_\mathrm{K}(t)=-\dd\ln\EEK/\dd\ln t$ and the decay time
$\tau_\mathrm{K}(t)=t/p_\mathrm{K}(t)$.
We relate $\tau_\mathrm{K}(t)$ to the turnover time
$\urms/\xiK$ through $\tau_\mathrm{K}=C_\mathrm{K}\xiK/\urms$,
and thus determine $C_\mathrm{K}$ as the late-time limit of
$C_\mathrm{K}(t)=[t/p_\mathrm{K}(t)]\,\urms(t)/\xiK(t)$, which
is defined analogously to \Eq{Ctau}.

\begin{figure}[t!]\begin{center}
\includegraphics[width=\columnwidth]{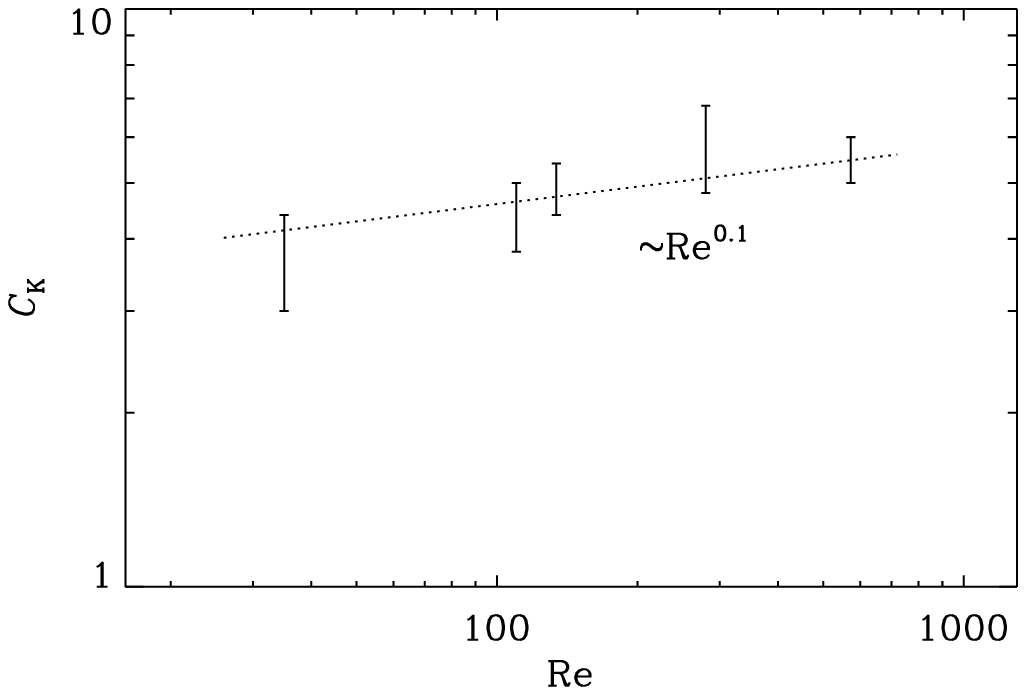}
\end{center}\caption[]{
Dependence of $C_\mathrm{K}$ on $\Rey$.
The line $\Rey^{0.1}$ is shown for comparison, but the data are also
nearly compatible with being independent of $\Rey$.
}\label{ptabH}\end{figure}

Purely hydrodynamic simulations can be performed by just ignoring the
magnetic field, or putting $\BB=0$.
In \Tab{TSummaryH} we summarize such simulations for different values
of $\nu$, which is quantified by the Reynolds number: $\Rey=\urms\xiK/\nu$.
\FFig{ptabH} shows, for $\Rey\ge100$, that $C_\mathrm{K}$ does not change
much with $\Rey$.
This was not expected.
It also confirms that the prolonged decay time found in
\Sec{ResistivityDependentDecay} is indeed a purely magnetic phenomenon.
Whether or not the resistivity dependence in the magnetic case must be
ascribed to reconnection remains an open question.
As discussed in \Sec{DecayTimes}, magnetic reconnection refers to
topologically constrained changes of magnetic field lines, but in the
present case of a turbulent magnetic field there is no connection with
the standard picture of reconnection.
An obvious alternative candidate for explaining the resistively prolonged
decay time may be related to magnetic helicity conservation in local
patches, as described by the conservation of the Hosking integral.
While this idea seems more plausible to us, it is not obvious how to
distinguish reconnection from magnetic helicity conservation in patches.
It is true that magnetic helicity would vanish in two dimensions,
but in that case it should be replaced by the anastrophy (see
\App{NoteAnastrophy} for a historical note on this word).
One aspect that might be different between the concepts of reconnection
and magnetic helicity conservation in patches could be the dependence on $\Pm$.
Our present results have not yet shown such a dependence, which might
support the idea that the dependence on resistivity is related to magnetic
helicity conservation in patches.

\begin{figure*}[t!]\begin{center}
\includegraphics[width=\textwidth]{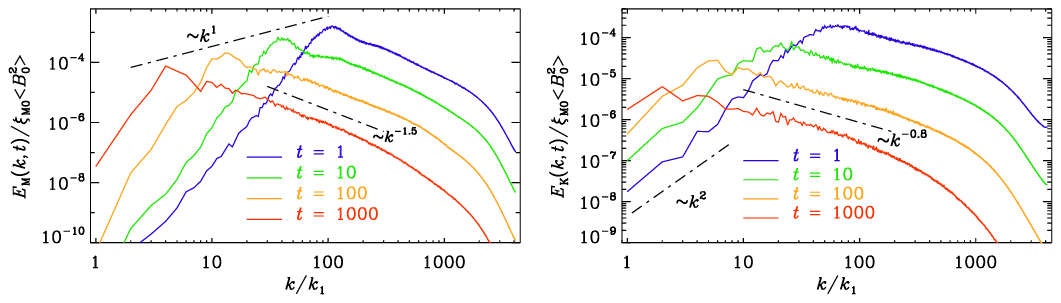}
\end{center}\caption[]{
Similar to \Fig{rspec_select_k60del2bc_k4_Pm5}, but for Run~2M1.
The magnetic peaks lie underneath a $k^\beta$ envelope
with $\beta=1$, as expected in the case of anastrophy conservation.
}\label{rspec_select_k200del2bc_k4_Pm5_8192c}\end{figure*}

\begin{figure*}[t!]\begin{center}
\includegraphics[width=\textwidth]{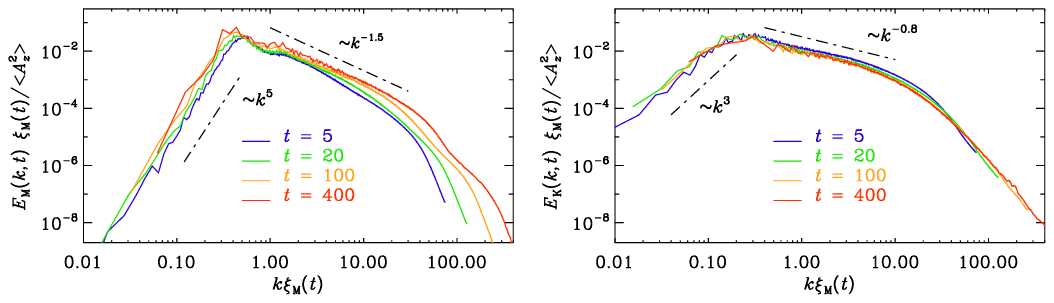}
\end{center}\caption[]{
Magnetic energy spectra (left) and kinetic energy spectra (right) for
Run~2m2 with $\Lu=1.8\times10^5$ and $\Pm=100$, at times $t=5$, 20, 100,
and 400, collapsed on top of each other by plotting both vs $k\xiM(t)$
and scaling them with $\xiM$.
This makes their heights agree, as expected.
}\label{rspec_compens_k200del2bc_k4_Pm100_16384a}\end{figure*}

\section{Hydromagnetic decay in two dimensions}
\label{2d}

We now perform 2D simulations where $\BB=\nab\times(\zzz A_z)$
lies entirely in the $xy$ plane (see \Tab{TSummary2D} for a summary).
\EEq{dAdt} then reduces to
\begin{equation}
\frac{\DD A_z}{\DD t}=\eta\nabla^2 A_z,
\label{dAdt2D}
\end{equation}
which obeys conservation of anastrophy, $\bra{A_z^2}=\const$
\citep{Fyfe+Montgomery76, Pouquet78, Pouquet93}, and the magnetic helicity
density vanishes pointwise, so the Hosking integral is then also zero.
These simulations are different from the recent ones by \cite{Dwivedi+24},
who performed 2.5D simulations.
In their case, there was a magnetic field component out of the plane.
The anastrophy was then not conserved and the Hosking integral was finite.

As we stated above, our present measurements of $C_\mathrm{M}$ as a
function of $\Lu$ cannot directly be compared with the reconnection rate
determined by \cite{Loureiro+12}, \cite{Comisso+15}, or \cite{Comisso+Bhattacharjee16}.
In addition, it is not obvious how relevant a 2D simulation is in
the present context because in 2D the anastrophy is conserved,
while the Hosking integral is strictly vanishing.

\begin{figure}[t!]\begin{center}
\includegraphics[width=\columnwidth]{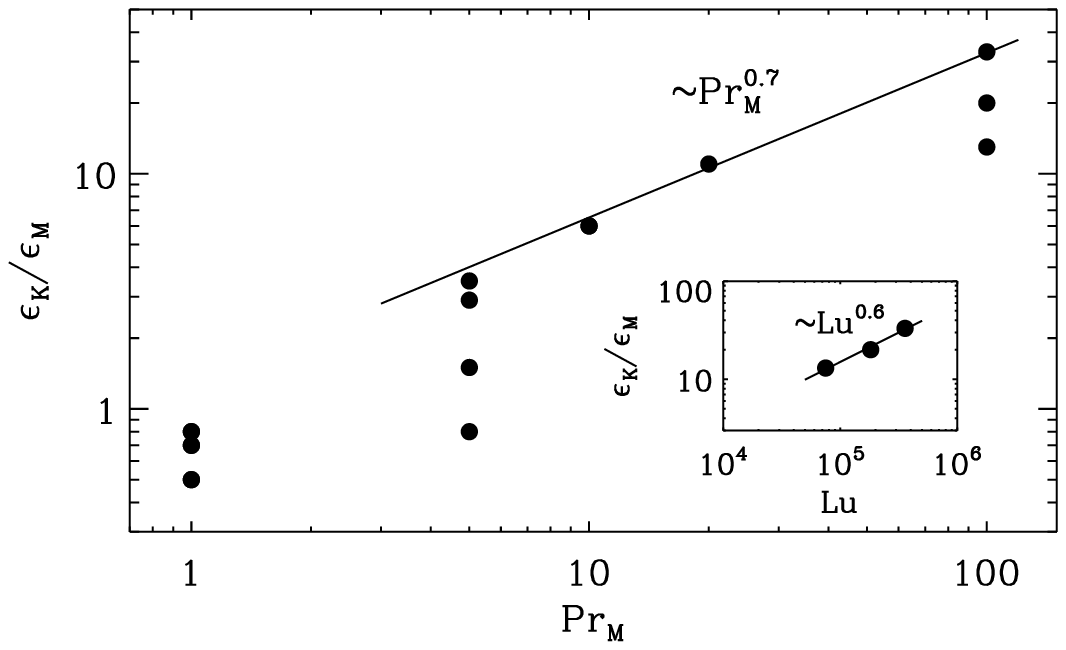}
\end{center}\caption[]{
Dependence of $\epsK/\epsM$ on $\Pm$.
The solid line denotes $1.3\,\Pm^{0.7}$, but many of the data points,
predominantly those with smaller $\Lu$, are beneath that line.
The inset shows that for $\Pm=100$, $\epsK/\epsM$ increases with $\Lu$
like $\Lu^{0.6}$.
}\label{ptab_pm_dep}\end{figure}

\begin{figure}[t!]\begin{center}
\includegraphics[width=\columnwidth]{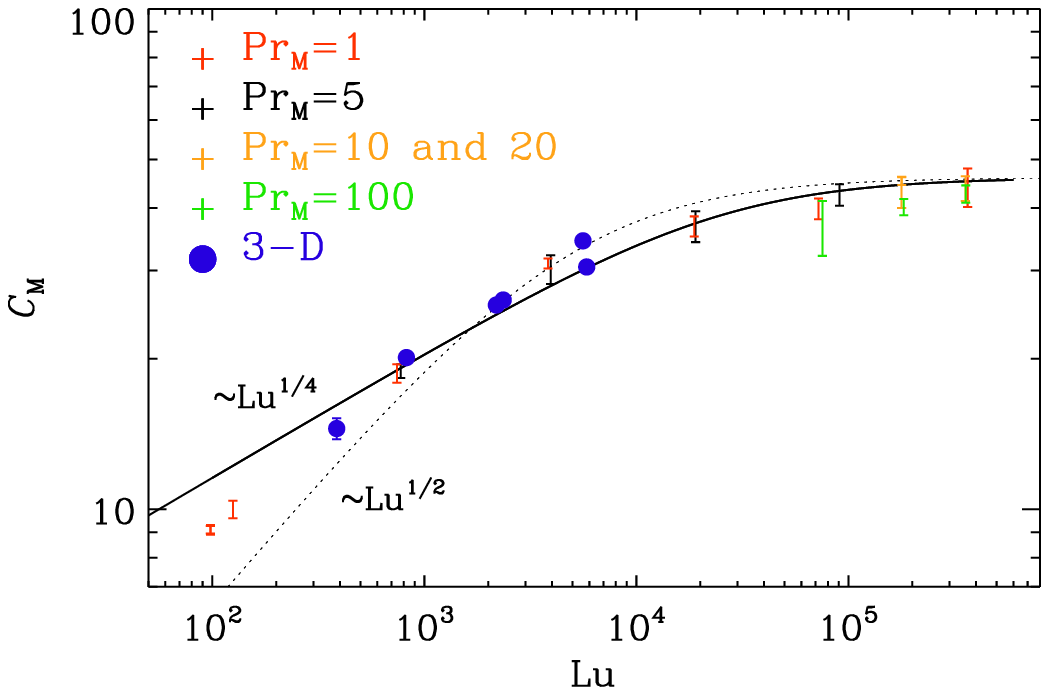}
\end{center}\caption[]{
Dependence of $C_\mathrm{M}$ on $\Lu$ for the 2D runs.
Shallow scaling $\propto\Lu^{0.1}$ is found for $10^4<\Lu<10^5$,
which is also compatible with a leveling off at $\Lu_\mathrm{c}=2.5\times10^4$,
as described by \Eq{LuFit} with $n=1/4$.
The black (red) data points are for $\Pm=5$ ($\Pm=1$).
The blue data points denote the 3D results from
\Sec{ResistivityDependentDecay}.
The orange symbols are for the runs with $\Pm=10$ and 20, listed in \Tab{TSummary2D}.
The thin dotted line gives \Eq{LuFit} with $n=1/4$ for comparison.
}\label{ptab2D}\end{figure}

\begin{figure*}[t!]\begin{center}
\includegraphics[width=\textwidth]{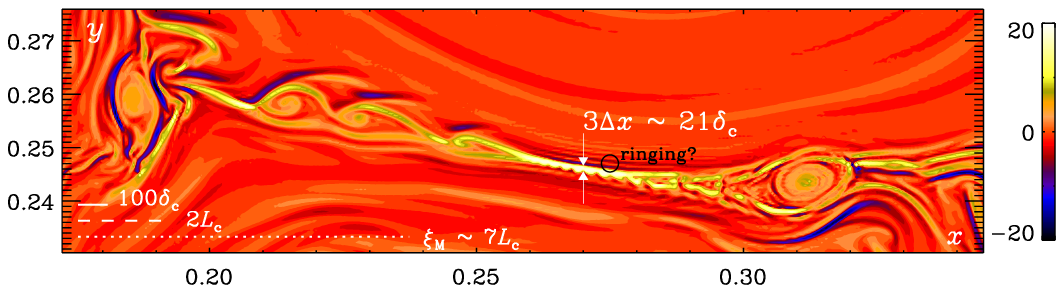}
\end{center}\caption[]{
Visualization of $J_z(x,y)$ of Run~2m6 with $\Pm=10$,
$\Lu=1.8\times10^5$, $\Lu_\nu\approx5\times10^4$, and $16384^2$
mesh points at $t=464$ for a small part of the domain with sizes
$2.8\xiM(t)\times0.74\xiM(t)$.
The lengths of $100\,\delta_\mathrm{c}$, $2L_\mathrm{c}$, and $\xiM$ are
indicated by horizontal white solid, dashed, and dotted lines, respectively.
The thickness of the current sheet corresponds to about
$3\Delta x\approx21\,\delta_\mathrm{c}$.
In its proximity, there are also indications of ringing, indicated
by the black circle.
}\label{ppjjs_last_k200del2bc_k4_Pm10_16384a}\end{figure*}

In our present purely 2D simulations, values of $\Lu$ up to about
$3\times10^5$ have been reached.
To allow for a longer nearly self-similar evolution, we used
$\kp/k_1=200$ instead of 60.
The simulation results give the prefactors in the scaling expected from
anastrophy conservation as
\begin{equation}
\xiM(t)\approx0.13\,\bra{A_z^2}^{1/4}\,t^{1/2},\quad
\EEM(t)\approx15\,\bra{A_z^2}^{1/2}\,t^{-1},
\label{xiMt2}
\end{equation}
where $\mu_0=\rho_0=1$ is used.
We also see from \Fig{rspec_select_k200del2bc_k4_Pm5_8192c}
that the spectral peaks evolve underneath an envelope
$\EM(k,t)\leq60\,\bra{A_z^2}\,k$.

\FFig{rspec_compens_k200del2bc_k4_Pm100_16384a} shows magnetic and
kinetic energy spectra for Run~2m2 with $\Lu=1.8\times10^5$ collapsed
on top of each other by plotting $\xiM^\beta(t)\EM\big(k\xiM(t)\big)$
versus $k\xiM(t)$ for $\beta=1$.
This plot suggests that the subinertial range scalings of $\EM(k,t)$
and $\EK(k,t)$ are proportional to $k^5$ and $k^3$, respectively.
Thus, they are steeper than expected in 3D.
This behavior is in some ways similar to the steepening observed for
helical decaying magnetic fields (see \citealt{BK17}).

In \Tab{TSummary2D} we also give the ratio $\epsK/\epsM$, which is seen
to increase with $\Pm$ (see \Fig{ptab_pm_dep}).
This was expected based on earlier results \citep{Bra14,Galishnikova+22},
but it was never shown in the decaying case of magnetically dominated
turbulence.
We see that $\epsK/\epsM\propto\Pm^{0.7}$, which is similar to the
previously studied case with large-scale dynamo action rather than the
case with just small-scale action where the slope was shallower.
The ratio $\epsK/\epsM$ is also found to increase with $\Lu$, at least
for $\Lu\ll10^6$.
As discussed in \cite{BR19}, an increase in $\epsK/\epsM$ with $\Pm$
may have implications for heating the solar corona, where the possible
dominance of kinetic energy dissipation over Joule dissipation is not
generally appreciated (see \citealt{Rappazzo+07, Rappazzo+18} for earlier
work discussing this ratio).

Although 2D and 3D runs are in many ways rather different from
each other, we now determine the same diagnostics as in the 3D case
(see \Tab{TSummary2D} and \Fig{ptab2D} for a plot of $C_\mathrm{M}$
versus $\Lu)$.
We see that the $C_\mathrm{M}$ dependence on $\Lu$ is qualitatively
similar for 2D and 3D turbulence.
Moreover, it becomes shallower for larger values of $\Lu$.
There is now evidence that $C_\mathrm{M}$ levels off and becomes
independent of $\Lu$.
It is possible to fit our data to a function of the form
\begin{equation}
C_\mathrm{M}(\Lu)\approx C_\mathrm{L}^{(1/4)}
\left(\frac{\Lu}{1+\Lu/\Lu_\mathrm{c}}\right)^{n},
\label{LuFit}
\end{equation}
where $\Lu_\mathrm{c}=2.5\times10^4$ is a critical Lundquist number
characterizing the point where the dependence levels off, $n=1/4$,
and $C_\mathrm{L}^{(1/4)}=3.7$.
The value of $\Lu_\mathrm{c}$ is larger than that found by
\cite{Loureiro+12}, where the asymptotes for small and large
Lundquist numbers cross at a value closer to 5000.
However, this difference could simply be related to different definitions
of the relevant length scales.
We note that our definition of $\xiM$ does not include a $2\pi$ factor.
A comparison with the Sweet--Parker value of $n=1/2$ results in reasonable
agreement for small values of $\Lu$, but there are rather noticeable
departures from the data for intermediate values.

Our 2D and 3D results for $C_\mathrm{M}$ are seen to be in good
agreement with each other, which is similar to the observation made
by \cite{Bhat+21}.
Obviously, larger simulations should still be performed to see whether
the agreement between 2D and 3D continues to larger Lundquist numbers.

It should also be noted that some of our data with their nominal error bars
do not lie on the fit given by \Eq{LuFit} with $n=1/4$.
Especially for $\Pm=100$ and smaller values of $\Lu$, we see that
$C_\mathrm{M}$ lies systematically below the fit.
It should be noted, however, that we would expect an increase with $\Pm$
as $(1+\Pm)^{1/2}$, which is the opposite trend, if the reconnection
phenomenology were applicable (see Eq.~(16) in \citealt{HS23}).

To check whether our simulations are sufficiently well resolved,
we show in \Fig{ppjjs_last_k200del2bc_k4_Pm10_16384a} a visualization of
$J_z(x,y)$ for Run~2m6 with $\Pm=10$ and $16384^2$ mesh points at $t=464$
for a small part of the domain with sizes $2.8\xiM(t)\times0.74\xiM(t)$
where a large current sheet breaks up into smaller plasmoids.
A comparison between Runs~2m5 and 2m6 with $8192^2$ and $16384^2$
mesh points is shown in \App{ResolvingPlasmoidInstability}.
Our main conclusions are that higher resolution suppresses the tendency to
produce ringing (i.e., the formation of oscillations on the grid scale),
but that the results for $C_\mathrm{M}$ are not very sensitive to the
numerical resolutions, as can be seen by comparing Runs~2m5 and 2m6
in \Tab{TSummary2D}.

\begin{figure*}[t!]\begin{center}
\includegraphics[width=\textwidth]{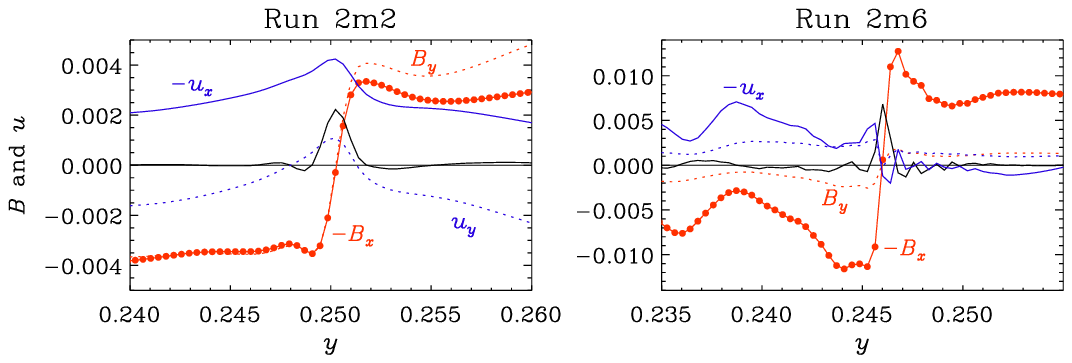}
\end{center}\caption[]{
Magnetic field profiles [$-B_x(y)$ and $B_y(y)$ in red]
and velocity profiles [$-u_x(y)$ and $u_y(y)$ in blue]
through the current sheet (in black) for Run~2m2
(through $x=0.372$) and Run~2m6 (through $x=0.270$).
}\label{pcurrent_sheet_comp}\end{figure*}

Next, we compare the typical length and thickness of current sheets in Run~2m6
with the values defined by \cite{Uzdensky+10} for critical current sheets,
$L_\mathrm{c}=\Lu_\mathrm{c}\,\eta/\vA$ and
$\delta_\mathrm{c}=L_\mathrm{c}/\Lu_\mathrm{c}^{1/2}$, respectively.
These lengths are indicated in \Fig{ppjjs_last_k200del2bc_k4_Pm10_16384a}.
Here we used $\Lu_\mathrm{c}=2.5\times10^4$ as the critical Lundquist
number, which was found to be representative of all of our cases.
We see that the current sheets in \Fig{ppjjs_last_k200del2bc_k4_Pm10_16384a}
have a length that is comparable to $\xiM\approx7\,L_\mathrm{c}$, because $\Lu\approx7\,\Lu_\mathrm{c}$.
The thickness of the current sheets is about $20\,\delta_\mathrm{c}$.
They are marginally resolved with about $3\Delta x$, where
$\Delta x=2\pi/16384\approx3.8\times10^{-4}$ is the mesh spacing.
Thus, the aspect ratio of thickness to length of the current
sheets is about $20\,\delta_\mathrm{c}/7\,L_\mathrm{c}=
(20/7)\,\Lu_\mathrm{c}^{-1/2}\approx0.02$.
This is about three times the nominal value estimated by
\cite{Uzdensky+10} for critical current sheets.

By comparison, \cite{HS23} estimated for the aspect ratio
$\delta_\mathrm{c}/\xiM=\Lu_\mathrm{c}^{1/2}/\Lu_\nu$.
For Run~2m6 with $\Lu_\nu\approx5\times10^4$, this yields 0.003,
which is about six times smaller than our value of 0.02.
While these discrepancies could perhaps be explained by the
absence of nondimensional prefactors in the definitions of
$\delta_\mathrm{c}$, we must also consider the possibility that
this is simply a consequence of a lack of resolution.

Runs~2m2 and 2m6 have in common that they have the same resistivity and
nearly the same value of $\Lu$ of about $1.8\times10^5$, but Run~2m2 has
a tenfold larger viscosity than Run~2m6, so $\Pm$ is increased from 10
to 100.
We see that $C_\mathrm{M}$ has deceased by only about 15\%, which is
much less than what is expected if $C_\mathrm{M}$ was proportional
to $\Pm^{-1/2}$.

In \Fig{pcurrent_sheet_comp} we show for Runs~2m2 and 2m6 magnetic
field profiles, $-B_x(y)$ and $B_y(y)$, and velocity profiles, $-u_x(y)$
and $u_y(y)$, through a particular current sheet.
We see that in Run~2m2 with $\Pm=100$, the profiles are much smoother,
even though the value of $\Lu=1.8\times10^5$ is the same in both cases.
Thus, the viscosity has a significant effect in smoothing the magnetic
field.
Even so, the effect on the value of $C_\mathrm{M}$ is small.

\section{Endpoints in the primordial evolutionary diagram}
\label{EndPoints}

The evolution of primordial magnetic fields is usually displayed in an
evolutionary diagram showing the comoving values of $\Brms$ versus $\xiM$
or, similarly, $\vA$ versus $\xiM$.
This dependence corresponds to a power law of the form $\vA=\vAz (\xiM/\xiMz)^\kappa$.
Since $\xiM(t)\sim t^q$, we have $\vA(t)\sim t^{-p/2}\sim\xiM^{-p/2q}$,
so $\kappa=p/2q=5/4$ for the Hosking scaling with $p=10/9$ and $q=4/9$.

Following \cite{BJ04}, the time $t$ in \Eq{tp_vs_xivA} would be replaced
by the age of the Universe at the time of recombination, $t_\mathrm{rec}$.
As emphasized by \cite{HS23}, this gives an implicit equation for the
magnetic field at recombination with the solution
$B(t_\mathrm{rec})\approx10^{-8.5}\G\,(\xiM/1\Mpc)=10^{-14.5}\G\,(\xiM/1\pc)$
if $C_\mathrm{M}=1$.
However, $B(t_\mathrm{rec})$ would be much larger and $\xiM$ much smaller
when $C_\mathrm{M}\gg1$ is taken into account.

Under the hypothesis of fast reconnection owing to plasmoid instability
\citep{Bhattacharjee+09, Uzdensky+10}, \cite{HS23} estimated
$C_\mathrm{M}$ as the square root of an effective cutoff value of about
$10^4$ for the Lundquist number and an additional factor of $\Pm^{1/2}$.
Here, they estimated $\Pm\approx10^7$, so $C_\mathrm{M}=10^{5.5}$
and $B(t_\mathrm{rec})\approx10^{-3}\G\,(\xiM/1\Mpc)=10^{-9}\G\,(\xiM/1\pc)$.

Our new results challenge the reliability of the anticipated dependence
of $C_\mathrm{M}$ on $\Pm$.
With the current results at hand, \Fig{ptab2D} suggests that $C_\mathrm{M}$
never exceeds the value $3.7\,\Lu_\mathrm{c}^{1/4}\approx47$, even for
$\Pm$ as large as 100.
Particularly important is the fact that this result is independent of $\Pm$.
Given that \cite{HS23} used $\Pm\approx10^7$, which yielded
an extra $10^{3.5}$ factor, and therefore $C_\mathrm{M}=10^{5.5}$ in
their estimate, our new findings imply that the resistivity effect
is actually independent of $\Pm$.
Although this has so far only been verified for values of $\Pm\leq5$,
we suggest that a more accurate formula for the endpoints of
the evolution with $C_\mathrm{M}=47\approx10^{1.7}$ would be
$B(t_\mathrm{rec})\approx10^{-6.8}\G\,(\xiM/1\Mpc)=10^{-12.8}\G\,(\xiM/1\pc)$.
These values are still above the lower limits inferred from suppressed
GeV photon emission from the halos of blazars \citep{Neronov+Vovk10}.

\section{Conclusions}
\label{Conclusions}

The present results have shown that, up to the largest Lundquist numbers
accessible to our present direct numerical simulations with $2048^3$
mesh points, the decay times depend on the resistivity.
Only for our 2D simulations do we see evidence for a cutoff.
The dependence of hydromagnetic turbulence properties on the value of
the resistivity is unusual for fully developed turbulence.
We wonder whether our results reflect just a peculiar property of
decaying turbulence or whether there could also exist aspects of
statistically stationary turbulence that depend on the microphysical
resistivity.
Possible examples of resistively controlled time dependences could
include the time that is needed to develop the final saturated magnetic
energy spectrum in kinetically forced turbulence, where the magnetic
field emerges due to a dynamo action \citep{HBD03, Scheko+04}.
A resistively slow adjustment phase is reminiscent of what occurs for
helical magnetic fields \citep{Bra01}, where it has also been possible
to measure a weak resistivity dependence of the turbulent magnetic
diffusivity \citep{Bran+08}.

In the present 3D case, the magnetic helicity vanishes on
the average.
However, in the spirit of the Hosking phenomenology of a decay
controlled by the conservation of the Hosking integral, it is very
possible, even in decaying turbulence, that the conservation of magnetic
helicity in patches of one sign of magnetic helicity plays an important
role in causing the resistively controlled decay speed.
Whether or not this is equivalent to talking about reconnection remains
an open question.
As discussed in \Sec{DecayTimes}, the idea of reconnection in terms of
current sheets and plasmoids may not be fully applicable in the context
of turbulence, where magnetic structures are more volume filling than
in standard reconnection experiments (see \Fig{pslice}).
Additional support for a possible mismatch between classical reconnection
theory and turbulent decay times comes from our numerical finding that
the dependence of $C_\mathrm{M}$ on $\Lu$ seems to be independent of
the value of $\Pm$ (see \Fig{ptab2D}).
More extensive numerical studies with resolutions of $8192^3$ meshpoints,
which was the resolution needed to see a leveling off in 2D,
might suffice to verify our 2D findings in the 3D case.

Our present work motivates possible avenues for future research.
One choice is to do the same for turbulence with a $-\alpha\uu$
friction term in the momentum equation.
Such calculations were already performed by \cite{BJ04} and the friction
term is also incorporated in the phenomenology of \cite{HS23}.
This is to model the drag from photons when their mean-free path begins
to exceed the scale $\xiM$.
This is the case after the time of recombination, when it contributes to dissipating
kinetic energy and leads to a decoupling of the magnetic field.
The field then becomes static and remains frozen into the plasma.
According to the work of \cite{HS23}, this results in a certain reduction
of $C_\mathrm{M}$ compared to the resistively limited value.
Verifying this with simulations would be particularly important. 

Another critical aspect to verify is the absence of a dependence of
$C_\mathrm{M}$ on $\Pm$ over a broader range of parameter combinations.
Given that there are always limitations on the resolution, it may be
useful to explore simulations in rectangular domains to cover a larger
range of scales.
Other possibilities include simulations with time-dependent values of
$\eta$ and $\nu$ to obtain a larger separation between $\xiM$ and the
dissipation scale near the end of the simulation.
However, there is the danger that artifacts are introduced that need
to be carefully examined.
Ill-understood artifacts can also be introduced by using hyperviscosity
and hyperresistivity, which are used in some simulations.
For the time being, however, the possibility of a $\Pm$ dependence of
$C_\mathrm{M}$ cannot be confirmed from our simulations.
Whether or not this automatically rules out reconnection as the reason
for a resistively limited value of $C_\mathrm{M}$, instead of the idea
of magnetic helicity conservation in smaller patches, remains uncertain.

As the value of $\Pm$ is increased, we also see a systematic increase
in the kinetic-to-magnetic dissipation ratio, $\epsK/\epsM$.
Such a dependence has previously been seen for kinetically dominated
forced turbulence, but it is shown here, perhaps for the first time,
for magnetically dominated decaying turbulence.
While such results may be of interest to the problem of coronal heating,
it should be remembered that the present simulations have large
plasma betas (i.e., the gas pressure dominates over the magnetic pressure).
It would therefore be of interest to check whether the obtained
$\Pm$-dependence also persists for smaller plasma beta.
The restriction to two dimensions is another computational simplification
that allowed significantly larger Lundquist numbers to be accessed, but it
needs to be checked that the results for $\epsK/\epsM$ are not very
sensitive to this restriction.
Comparing the 3D Runs~M1 and M3 of \Tab{TSummary} with the 2D Run~2M3 of
\Tab{TSummary2D}, which have $\Pm=5$ and similar Lundquist numbers,
we see that the 2D results may overestimate the ratio $\epsK/\epsM$
by a factor of about two.
Future work will need to show whether this can also be confirmed for
other parameters.

\begin{acknowledgements}
We thank Pallavi Bhat (Bangalore) and
David Hosking (Princeton) for detailed comments and suggestions,
especially regarding reconnection in the 2D case.
We are also acknowledge inspiring discussions with the participants of
the program on ``Turbulence in Astrophysical Environments'' at the Kavli
Institute for Theoretical Physics in Santa Barbara.
We are also grateful to S\'ebastien Galtier, Romain Meyrand, Annick Pouquet,
and Alex Schekochihin for their interest in shedding light on the origin
of the word anastrophy; see \App{NoteAnastrophy}.
This research was supported in part by the 
Swedish Research Council (Vetenskapsr{\aa}det) under Grant No.\ 2019-04234,
the National Science Foundation under Grant No.\ NSF PHY-1748958,
a NASA ATP Award 80NSSC22K0825, and
the Munich Institute for Astro-, Particle and BioPhysics (MIAPbP), which
is funded by the Deutsche Forschungsgemeinschaft (DFG, German Research
Foundation) under Germany´s Excellence Strategy - EXC-2094 - 390783311.
We acknowledge the allocation of computing resources provided by the
Swedish National Allocations Committee at the Center for
Parallel Computers at the Royal Institute of Technology in Stockholm. 
F.V.\ acknowledges partial financial support from  the Cariplo ``BREAKTHRU'' funds Rif: 2022-2088 CUP J33C22004310003.

\vspace{2mm}\noindent
{\em Software and Data Availability.} The source code used for
the simulations of this study, the {\sc Pencil Code} \citep{JOSS},
is freely available on \url{https://github.com/pencil-code/}.
The DOI of the code is https://doi.org/10.5281/zenodo.2315093.
The simulation setups and corresponding data are freely available on
\url{https://doi.org/10.5531/sd.astro.9} \citep{DATA} and
\url{https://doi.org/10.5281/zenodo.10527437}.
\end{acknowledgements}

\bibliography{ref}{}
\bibliographystyle{aa}

\clearpage

\appendix
\section{Historical note on anastrophy}
\label{NoteAnastrophy}

In recent years, the term anastrophy for the mean squared magnetic vector
potential $\bra{A_z^2}=\const$ has become increasingly popular
\citep{Tronko+13, Galtier+Meyrand15, Zhou+21, HS21, Scheko22}.
In the 1970s, it was referred to as mean square vector potential
\citep{Fyfe+Montgomery76} or as the variance of the magnetic potential
\citep{Pouquet78}.
The term anastrophy was first used in the 1987 Les Houches lecture notes
by \cite{Pouquet93}, and it was also used by \cite{Vakoulenko93}, but
neither of them provided an explanation of its origin.

Annick Pouquet (2024, private communication) informed us now that the word
may have been invented by Uriel Frisch and Nicolas Papanicolaou during
a meeting on a Winter Sunday in the late 1970s at Saint-Jean-Cap-Ferrat,
while she and Jacques L\'eorat were also present.

The word has Greek roots, where `strophe' refers to curl or turning,
and `ana' therefore hints at the inverted curl of $\BB$.
This is somewhat reminiscent of the word palinstrophy, which is the
mean squared double curl of the velocity, where `palin' means again.
This term was also invented by Frisch and Papanicolaou.
The palinstrophy is proportional to the rate of change of enstrophy,
i.e., the mean squared vorticity.

\section{Resolving the plasmoid instability}
\label{ResolvingPlasmoidInstability}

\begin{figure*}[t!]\begin{center}
\includegraphics[width=.49\textwidth]{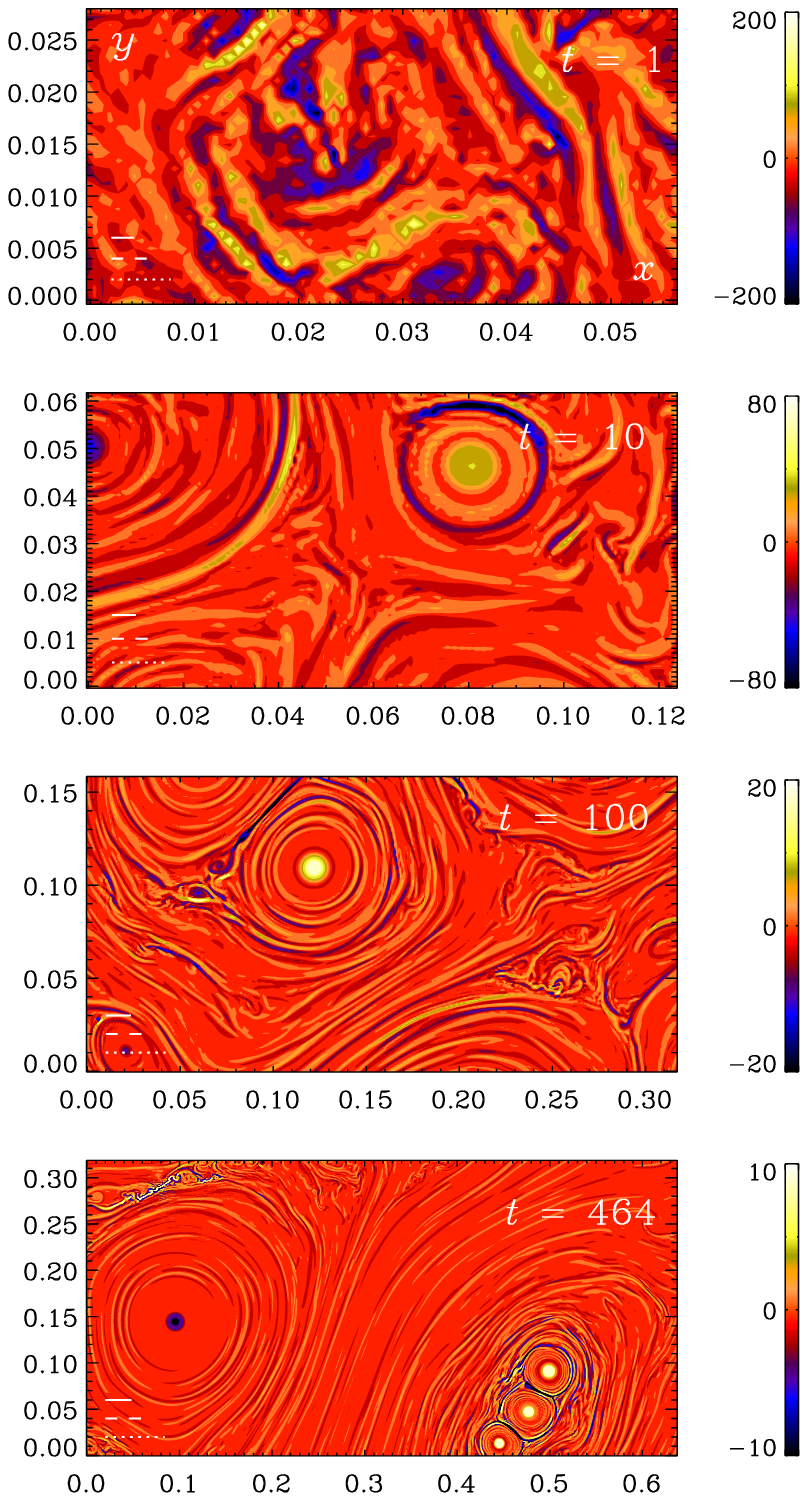}
\includegraphics[width=.49\textwidth]{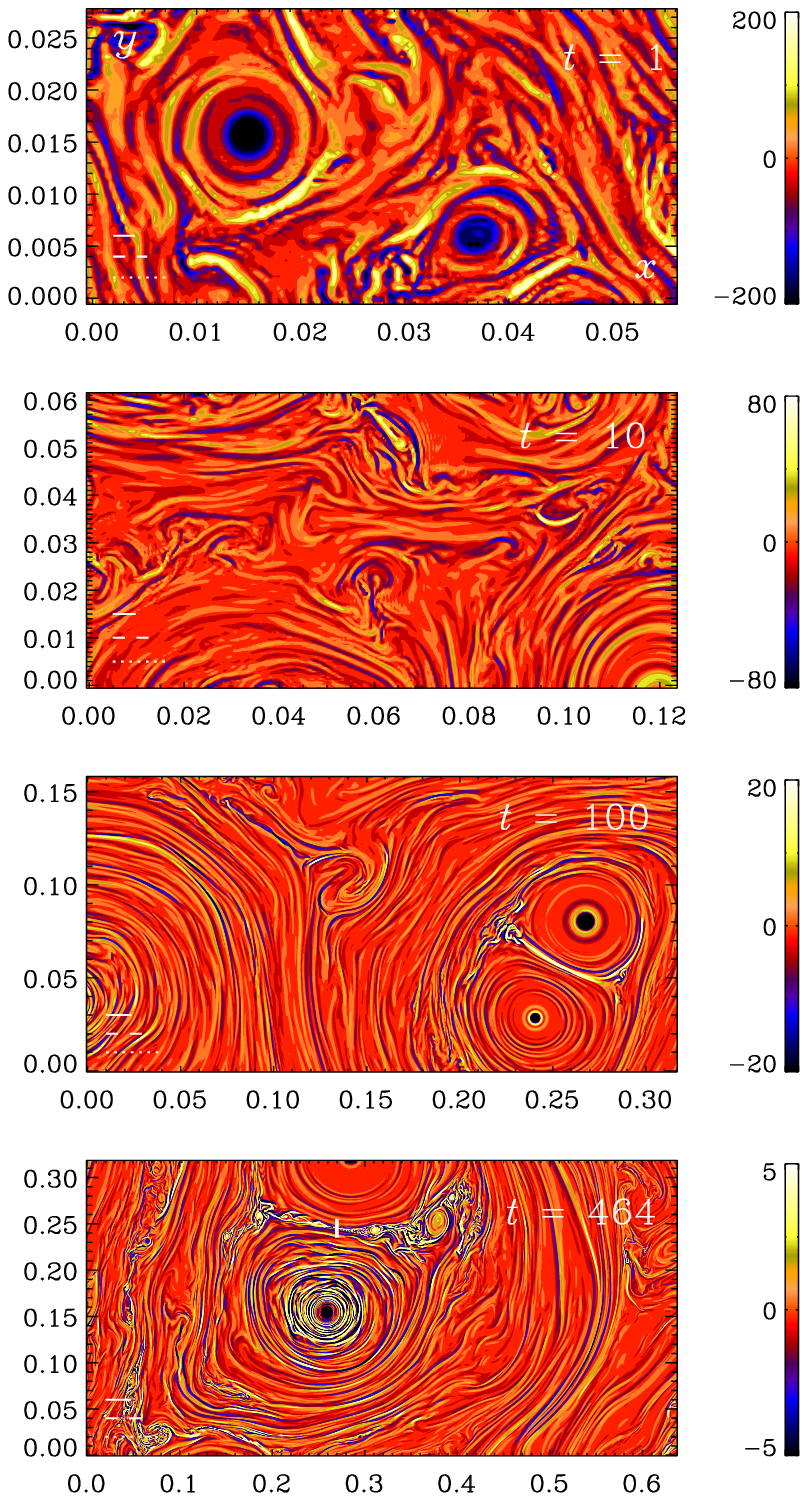}
\end{center}\caption[]{
Comparisons of visualizations of $J_z(x,y)$ for Runs~2m5 and 2m6 with $\Pm=10$
and $8192^2$ (left) and $16384^2$ (right) mesh points at times $t=1$, 10, 100, and 464.
In each panel the lengths of the dotted, dashed, and solid lines denote
the values of $\xiM$, $5\,L_\mathrm{c}$, and $500\,\delta_\mathrm{c}$.
The bottom right panel shows the same current sheet that was
presented in \Fig{ppjjs_last_k200del2bc_k4_Pm10_16384a} as a blow-up.
The thick white line in that panel at $(x,y)=(0.27,0.25)$ marks the
location of the cross section shown in \Fig{pcurrent_sheet_comp}.
}\label{ppjjs_k200del2bc_k4_Pm10_16384a}\end{figure*}

\begin{figure*}[t!]\begin{center}
\includegraphics[width=.49\textwidth]{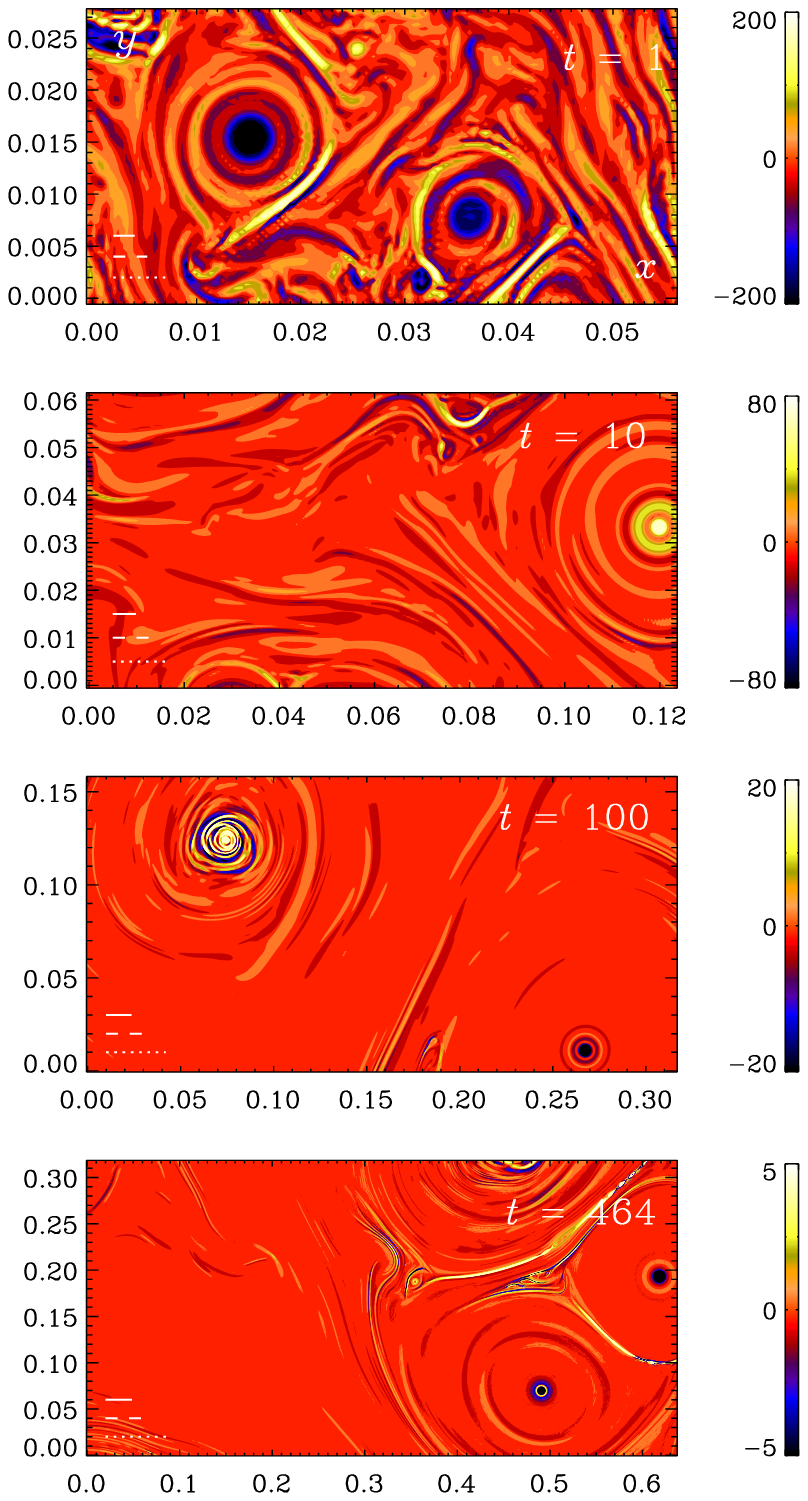}
\includegraphics[width=.49\textwidth]{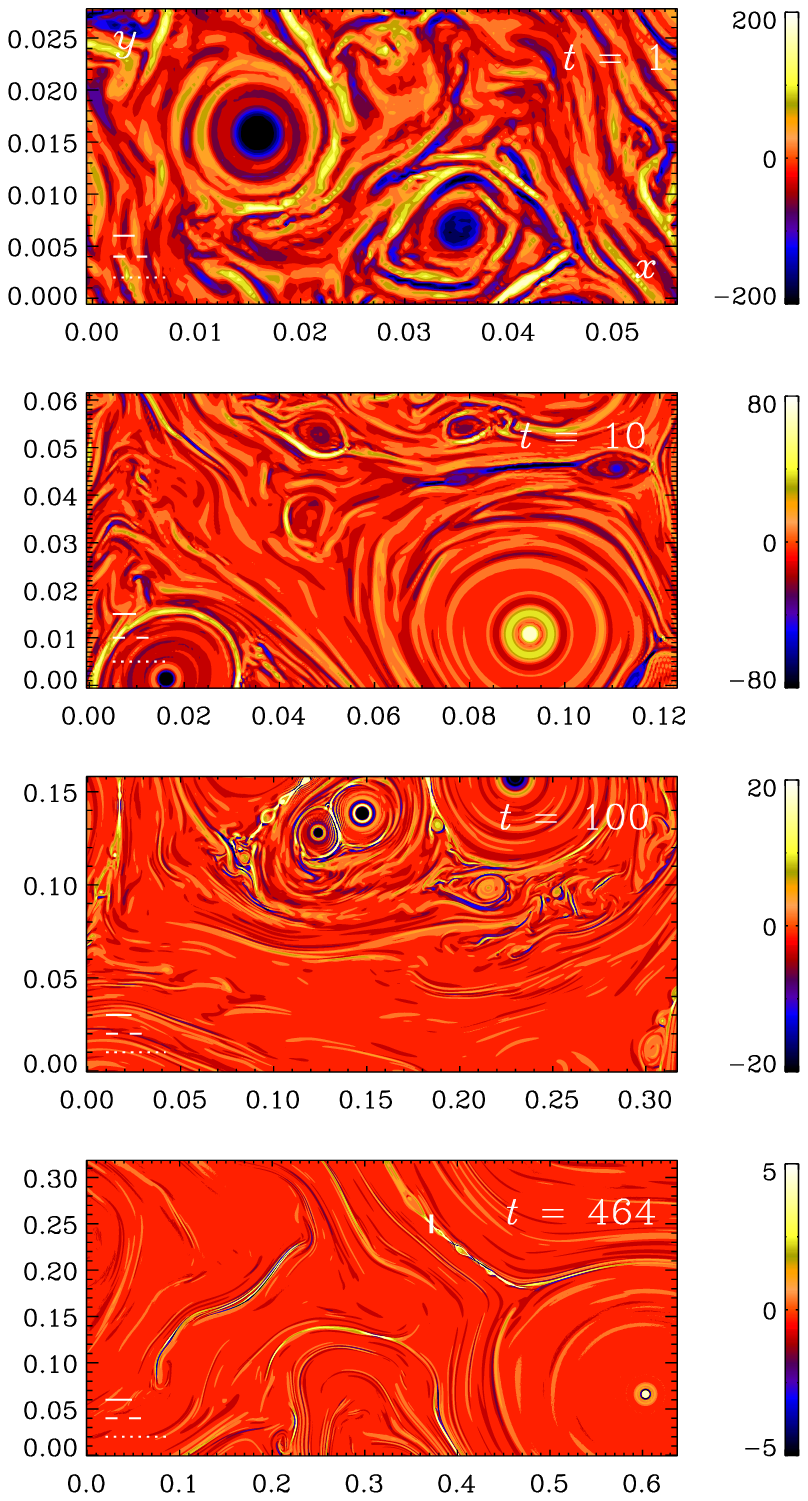}
\end{center}\caption[]{
Similar to \Fig{ppjjs_k200del2bc_k4_Pm10_16384a}, but
for Run~2m1 with $\Lu=75,000$ (left) and Run~2m2 with $\Lu=182,000$ (right)
using $\Pm=100$ and $16384^2$ mesh points in both cases.
The thick white line in the bottom right panel at $(x,y)=(0.37,0.25)$ marks
the location of the cross section shown in \Fig{pcurrent_sheet_comp}.
}\label{ppjjs_k200del2bc_k4_Pm100_16384b}\end{figure*}

The expected thickness of critical layers (i.e., the minimum thickness of
current sheets when they become unstable to the plasmoid instability)
is expected to be on the order of $\delta_\mathrm{c}$ (see \Sec{2d}).
To compare this with our 2D simulations, we show in
\Fig{rspec_select_k200del2bc_k4_Pm5_8192c} visualizations of $J_z(x,y)$
in parts of the full domain with sizes $5\xiM\times10\xiM$ at times $t=1$,
10, 100, and 464.
We note that $\xiM(t)$ increases with time, so the $x$ and $y$ ranges
increase with time too.
We see that each panel contains about one pair of current tubes.
Furthermore, the other length scales, $L_\mathrm{c}$ and
$\delta_\mathrm{c}$, increase with time by the same factor, as expected
for a self-similar evolution.

At late times, for $t=100$ and $464$, the current sheets are seen to break
up into plasmoids.
They are obviously better resolved in Run~2m6 with $16384^2$ mesh points
than in Run~2m5 with only $8192^2$ mesh points.
At lower resolution, there is  a higher tendency for ringing (i.e., the
formation of oscillations on the grid scale), which indicates that the
resolution limit has been reached.
Nevertheless, \Tab{TSummary2D} and \Fig{ptab2D} show that, within the error
bars, the values of $C_\mathrm{M}$ are similar for the two resolutions.

It is possible that the critical current sheets are not well resolved
during a significant fraction of the duration of the simulation.
Theoretically, we expect the thinnest sheets to set the reconnection
rate, but this is also the place where the value of $\Pm$ matters because
viscosity is unimportant for the larger scales in the plasmoid hierarchy.
Therefore, if we do not resolve that sheet, it seems reasonable that
we would also not see a dependence on $\Pm$.
To check this, we now consider a version of Run~2m2, where the viscosity
is ten times larger, which increases $\Pm$ from 10 to 100; see Run~2m6.
The result is shown in \Fig{ppjjs_k200del2bc_k4_Pm10_16384a}.
We see that a lower value of $\Lu$ for $\Pm=100$ changes the results
in an expected fashion, thus rejecting the possibility that the
weak dependence of $C_\mathrm{M}$ on $\Pm$ was an artifact of having
chosen unreliably large values of $\Lu$; compare Runs~2m1 and 2m2 in
\Fig{ppjjs_k200del2bc_k4_Pm100_16384b}.

In the bottom right panel of \Fig{ppjjs_k200del2bc_k4_Pm10_16384a},
we see the same current sheet that was already presented in \Sec{2d} as
\Fig{ppjjs_last_k200del2bc_k4_Pm10_16384a}.
However, we also see in \Fig{ppjjs_k200del2bc_k4_Pm10_16384a} that
there are many other current sheets that are not yet in the process of
breaking up.

\end{document}